\documentclass[english,journal=jpcafh,etalmode=truncate,maxauthors=90,manuscript=article, layout=twocolumn]{achemso}
\setkeys{acs}{usetitle=true,etalmode=truncate,maxauthors=90}

\usepackage{natbib}
\usepackage{mciteplus}
\usepackage{natmove}

\author{Miwa Goto}
\affiliation[Universit\"ats-Sternwarte M\"unchen]
{Universit\"ats-Sternwarte M\"unchen, Scheinerstr. 1, D-81679
  Munich, Germany}
\email{mgoto@usm.lmu.de}

\author{Nick Indriolo} 
\affiliation[Johns Hopkins University]{Department of Physics and
  Astronomy, 3400 N. Charles St., Baltimore, MD 21218, USA}

\author{T. R. Geballe}
\affiliation[Gemini Observatory]{670 North A`ohoku Place, Hilo,
  HI 96720, USA}

\author{T. Usuda} 
\affiliation[Subaru Telescope]{650 North A`ohoku Place, Hilo, HI
  96720, USA}

\title[H$_3^+$ in the Central Few Parsecs of the Galaxy]
  {H$_3^+$ Spectroscopy and the Ionization Rate of Molecular
    Hydrogen in the Central Few Parsecs of the Galaxy \footnote{
Based on data collected in the course of CRIRES
  Science Verification program (60.7A-9057) and open-use program (079.C-0874)
  at the  VLT on Cerro Paranal (Chile), which is operated
  by the European Southern Observatory (ESO). Based also on data
  collected at Subaru Telescope, which is operated by the
  National Astronomical Observatory of Japan. }}

\begin{document}
\begin{abstract}

We report observations and analysis of infrared spectra of
H$_3^+$ and CO lines in the Galactic center, within a few
parsecs of the central black hole, Sgr A*.  We find a cosmic ray
ionization rate typically an order of magnitude higher than
outside the Galactic center. Notwithstanding, the elevated
cosmic ray ionization rate is 4 orders of magnitude too short to
match the proton energy spectrum as inferred from the recent
discovery of the TeV $\gamma$-ray source in the vicinity of
Sgr~A*.

\end{abstract}

\section{Introduction}

\subsection{H$_3^+$ and the cosmic ray ionization rate}

The ionization of molecular hydrogen by cosmic rays -- mainly
high energy protons, helium nuclei, and electrons -- has diverse
and important influences on the physics and chemistry of
interstellar molecular clouds. It is a significant heat source
in interstellar clouds, through the secondary electrons
liberated in the ionization process. If a cloud is even slightly
ionized, its internal motions are restricted by the ambient
magnetic field. In star forming clouds the timescale for cloud
collapse is affected by the ionization fraction. Ion-neutral
reactions, which generally proceed with large Langevin rates,
are the main propellant of interstellar chemistry;
neutral-neutral reactions have reaction barriers and are
consequently prohibitively slow at the low-temperatures of
interstellar molecular clouds. Thus chemistry in molecular
clouds is driven by the cosmic ray ionization of H$_2$.

Interstellar molecular clouds come in a range of sizes and
densities, but for many purposes may be regarded as consisting
of two types: diffuse clouds (10~cm$^{-3}$ $<$ $n$ $<$
3~$\times$~10$^2$~cm$^{-3}$) and dense clouds ($n$ $>$
3~$\times$~10$^2$~cm$^{-3}$). Regardless of the cloud type the
ionization of H$_2$ within a cloud occurs almost exclusively by
cosmic rays collisions. Photons capable of ionizing H$_2$ (i.e.,
with $E>$15.4~eV)\cite{Herzberg:1969p37807} are consumed outside
the cloud or on its surface by the ionization of atomic hydrogen
(ionization potential 13.6~eV) and/or absorption by dust
particles and thus have little or no effect on physics and
chemistry in the interior. Ultraviolet photons with less than
this energy penetrate diffuse clouds; as a result hydrogen in
them is only partly in molecular form and the abundant element
carbon is mainly atomic and singly ionized due to its low
ionization potential (11.3~eV). In dense clouds dust does not
allow near-ultraviolet radiation to penetrate. As a result dense
clouds are essentially fully molecular (e.g., virtually all H in
H$_2$ and all C in CO).

The ionization collisional cross-section of H$_2$ peaks at
70~keV \cite{Rudd:1983p36770,Padovani:2009p31095} and decreases
to higher energy as a power law. It is thought that the cosmic
rays that contribute most significantly to the ionization of
H$_2$ within clouds have energies, $E<100$~MeV. Their intrinsic
spectrum, however, cannot be directly measured from inside the
solar system, because the influx of cosmic rays with $E<1$~GeV
is deflected by the solar wind and magnetic field. One must
therefore rely on indirect techniques to determine  the
  cosmic ray ionization rate of molecular hydrogen, $\zeta_2$,
either by appealing to the energetics of the interstellar medium
\cite{Spitzer:1968p31337}, or to the chemistry by observing the
abundances of molecules formed in reactions triggered by cosmic
ray ionization of H$_2$, such as OH, HD, HCO$^+$
\cite{Black:1977p37990,vanDishoeck:1986p37944,vanderTak:2000p31226},
or H$_3^+$.

Of these molecules H$_3^+$, first detected in molecular clouds
17 years ago \cite{Geballe:1996p8673}, is the most reliable
probe of the cosmic ray ionization rate of molecular hydrogen,
simply because the number of the reactions involved in its
production is effectively the minimum, one. In a molecular
cloud, once H$_2$ is ionized, the reaction, H$_2^+$ + H$_2$
$\rightarrow$ H$_3^+$ + H, is so rapid compared to the other
competing processes that virtually all H$_2^+$ produced by
cosmic rays quickly is converted into H$_3^+$. The H$_3^+$
formation rate is therefore directly proportional to the cosmic
ray ionization rate $\zeta_2$. Destruction of H$_3^+$ is
dominated by fast electron dissociative recombination in diffuse
clouds and by chemical reaction with CO to form HCO$^+$ in dense
clouds. Both reactions have been well studied in the
laboratory. Equating creation and destruction rates one obtains
(1)
\[n({\rm H_2}) \, \zeta_2 = k_{\rm CO} \, n({\rm H_3^+})\,n({\rm CO})\] 
in dense clouds and (2) 
\[n({\rm H_2}) \, \zeta_2 = k_{\rm e} \, n({\rm
  H_3^+})\,n({\rm e})\] in diffuse clouds, where $k_{\rm e}$ and
$k_{\rm CO}$ are the dominant H$_3^+$ destruction rate
coefficients for the two types of clouds, and $k_{\rm e}$ is
larger than $k_{\rm CO}$ by roughly two orders of magnitude at
typical cloud temperatures. In each type of cloud, however, the
steady state density of H$_3^+$ depends linearly on the cosmic
ray ionization rate.

\subsection{High local cosmic ray ionization rate}

Until a decade ago, $\zeta_2$ was simplistically considered to
be roughly uniform throughout much of the Galaxy, with values
within a factor of 3 of 3$\times$10$^{-17}$~s$^{-1}$, because
the great penetrating power of typical cosmic rays was thought
to smooth out local influences of particle accelerators. Studies
of interstellar clouds using H$_3^+$ have now conclusively
demonstrated that this is not the case. They have revealed that
values of $\zeta_2$ in diffuse clouds are on average an order of
magnitude higher than in dense molecular clouds
\cite{McCall:2002p8292,Indriolo:2007p8061,Indriolo:2012p31279},
a finding most easily explained by a large and heretofore
unrecognized population of low energy ($<$~10 MeV) cosmic rays,
which penetrate diffuse clouds much more effectively than dense
clouds \cite{Indriolo:2009p8835}. In addition, real and large
differences in the rates deduced for different diffuse clouds
have been found \cite{Indriolo:2007p8061}. Finally, a direct
connection between a particle acceleration source and the local
cosmic ray ionization rate has been demonstrated by
\citeauthor{Indriolo:2010p31286}\cite{Indriolo:2010p31286} near
the supernova remnant IC~443, where $\zeta_2$ = $2\times
10^{-15}$~s$^{-1}$, an order of magnitude higher than in typical
diffuse clouds in the Galaxy.

Much higher ionization rates than those reported above are
predicted in especially energetic environments, within galactic
nuclei or near the highest energy supernovae.
\citeauthor{YusefZadeh:2007p7218}\cite{YusefZadeh:2007p7218} has
argued that the ionization rate could be as high as $5\times
10^{-13}$~s$^{-1}$ near the center of the Milky Way galaxy, in
locations where high energy electrons that produce X-ray and
non-thermal radio emission encounter dense molecular
clouds. Likewise,
\citeauthor{Tatischeff:2012p41073}\cite{Tatischeff:2012p41073}
estimate that the ionization rate in the Galactic center's
Arches Cluster is $1\times 10^{-13}$~s$^{-1}$. The prediction by
\citeauthor{Becker:2011p31110}\cite{Becker:2011p31110} (also
\citeauthor{Black:2012p37714}\cite{Black:2012p37714}) is even
more extreme: ionization rates of $\sim$$1\times
10^{-12}$~s$^{-1}$ near $\gamma$-ray emitting supernova
remnants, if they are the principal sources of Galactic cosmic
rays.

 The subject of this paper is the ionization rate in what
  may be the closest dense molecular cloud to the Galactic
  center where H$_3^+$ $R$(2,2)$^l$ absorption was first
  detected in the interstellar medium by
  \citeauthor{Goto:2008p981}
  (\citeyear{Goto:2008p981})\cite{Goto:2008p981}. The ionization
  rate measured using the new data with improved velocity
  resolution is compared to the ionization rates expected from
  the X-ray and the cosmic ray flux in the central few parsecs
  of the Galactic center.


\subsection{The Central Molecular Zone} 

As suggested above, the very innermost region of the Galaxy is a
site where one might expect to find extreme values of
$\zeta_2$. 
The rate per
unit volume of supernova outbursts in the center is estimated to
be 0.04 times per century \cite{Crocker:2011p35207}, 2000 times
higher than the Milky Way average
\cite{Tammann:1994p40281,vandenBergh:1991p40259}. Despite these
and other violent events and the generally high energy density,
molecular gas is abundant in the center. Indeed the central
200~pc of the Galaxy, known as the Central Molecular Zone
\cite{Morris:1996p19505}, contains one tenth of the entire
molecular mass of the Milky Way in its scant 0.001\% of the
volume \cite{Morris:1996p19505}.

A significant fraction of the molecular gas in the Central
Molecular Zone is found in giant and dense molecular clouds,
which take up only a small fraction of the volume of the Central
Molecular Zone. However, spectroscopy of H$_3^+$ has recently
shown that molecules also are plentiful outside of those
clouds. Indeed a large fraction of the volume of the Central
Molecular Zone is filled with warm ($\sim$250~K) and much more
rarified ($\le$100~cm$^{-3}$) molecular gas
\cite{Goto:2002p1051,Oka:2005p8161}. Within this diffuse
molecular gas $\zeta_2$ has been estimated to be
(2--7)~$\times$~10$^{-15}$~s$^{-1}$ \cite{Oka:2005p8161},
roughly an order of magnitude higher than the average value for
diffuse clouds outside of the center.

\begin{figure}[bth]
\begin{center}
\hspace*{-5mm}
\includegraphics[width=0.54\textwidth,angle=0]{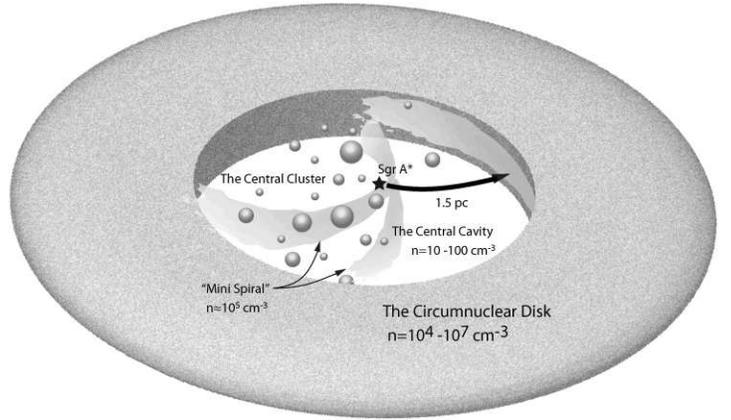}
\end{center}
  \caption{Schematic view of the central few parsecs of the Galaxy, 
showing the central black hole (Sgr A*), members of the Central Cluster 
of massive stars, the Circumnuclear disk, and molecular gas within the 
Central Cavity.}
\label{fgr:schematic}
\end{figure}

\subsection{The central few parsecs}

Here we focus on molecular gas within the central few parsecs of
the Central Molecular Zone, a tiny region containing a massive
black hole Sgr~A*\cite{Gillessen:2009p6070}, a multitude of old
stars, and a "Central Cluster" of $\sim$10$^2$ young, hot and
massive stars (also known as the nuclear star
  cluster \cite{Krabbe:1995p7344}). A sketch of this
"laboratory" is shown in \ref{fgr:schematic}. The principal
gaseous structure there is the Circumnuclear Disk
\cite{Christopher:2005p7283,MonteroCastano:2009p6848}, a clumpy
stream of molecular clouds, with inner radius 1.5 pc, orbiting
Sgr~A*. The gas density drops from 10$^4$--10$^7$~cm$^{-3}$
within the Disk
\cite{Gusten:1987p19488,RequenaTorres:2012p40543} to
10--100~cm$^{-3}$ in the Central Cavity
\cite{Baganoff:2003p36758}, the low-density region between
Sgr~A* and the inner surface of the Circumnuclear Disk. Gas in
the Central Cavity is partly ionized by the radiation from the
stars in the Central Cluster
\cite{Becklin:1968p40541,Viehmann:2005p5622}. The high-mass
stars in the Central Cluster formed approximately 6~Myrs ago,
and are now at the end of the main-sequence evolutionary phase
\cite{Paumard:2006p5913}. Among them, two of the brightest at
mid-infrared wavelengths are GCIRS~1W and GCIRS~3. Gas in the
cavity is partly ionized by the hot stars in the Central
Cluster. The radiation and the stellar winds from those stars
are the main source of radiation and kinetic energy injected
into the Central Cavity
\cite{Najarro:1997p6043,Martins:2007p7448}. The HII region of
the Central Cavity together with a few distinct streamers of the
molecular clouds (``Mini-spiral'') are collectively called
Sgr~A~West \cite{Zhao:2009p5799,Zhao:2010p40563}.

\section{Observations}

Absorption spectroscopy in astronomy is similar to absorption
spectroscopy in the laboratory. In each case the experimental
setup consists of a background light source, the spectrograph,
and a sample placed between the two. For this study the light
sources were GCIRS~3 and GCIRS~1W in the Central Cluster within
a few tenths of a parsec of Sgr~A* as viewed in the plane of the
sky. The spectrograph was the Cryogenic Infrared Echelle
Spectrograph (CRIRES) \cite{Kaeufl:2004p40429} ($\lambda/\Delta
\lambda =$50,000--100,000) on the Very Large Telescope (VLT) at
the Paranal Observatory in Chile. The sample was the gas between
the Earth and the Galactic center, which are separated by
8~kpc\cite{Eisenhauer:2005p6146,Ghez:2008p41514}. In actuality
almost all of the intervening gas is within the Central
Molecular Zone or associated with Galactic spiral arms between
the Central Molecular Zone and the Earth.

The targeted spectral lines of H$_3^+$ were v$_2$ vibrational
transitions from the ($J$,$K$)=(1,1), (3,3) and (2,2) levels at
$\lambda=$3.5--3.7~$\mu$m. Each of these lines is an important
diagnostic. The line from the (1,1) level provides in a
straightforward manner the column density in the lowest energy
level. The (3,3) level, 361~K above (1,1) is metastable; when
collisionally populated, as is the case in the Central Molecular
Zone, it serves as a (density-independent) thermometer. The
(2,2) level, at an intermediate energy, has a lifetime against
spontaneous emission of 27 days; for temperatures not much lower
than its excitation, 150~K, lines from (2,2) are densitometers
for low density clouds. In addition to these lines of H$_3^+$,
the fundamental band of $^{13}$CO at 4.7~$\mu$m was observed, in
part to help distinguish between gas in the Central Molecular
Zone and the foreground arms. (Lines of $^{12}$CO are also
present but are too strong to be useful in this regard.)

The data were obtained on several nights in 2006--2008. Standard
data reduction techniques were employed (e.g,
\citeauthor{Goto:2008p981} \citeyear{Goto:2008p981}
\cite{Goto:2008p981}). The final spectra, shown in Figs. 2-4,
are corrected for instrumental transmission and atmospheric
absorption and have been wavelength-calibrated to an accuracy
corresponding to $\pm$1~km\,s$^{-1}$.

\begin{figure}[bth]
\begin{center}
\hspace*{-12mm}
\includegraphics[height=0.48\textwidth,angle=-90]{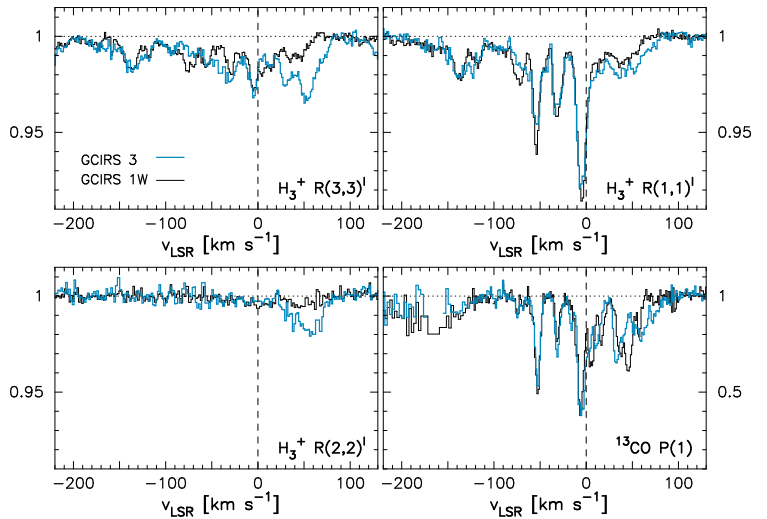}
\end{center}
 \vspace{10mm}
 \caption{H$_3^+$ and $^{13}$CO~v=1-0 $P$(1) spectra toward
   GCIRS~3 (blue) and GCIRS~1W (black).}
   \label{fgr:comparison}
\end{figure}

\section{Results}

\subsection{Gas in the Central Molecular Zone}

\begin{figure}[bth]
\begin{center}
\includegraphics[height=0.53\textheight,angle=0]{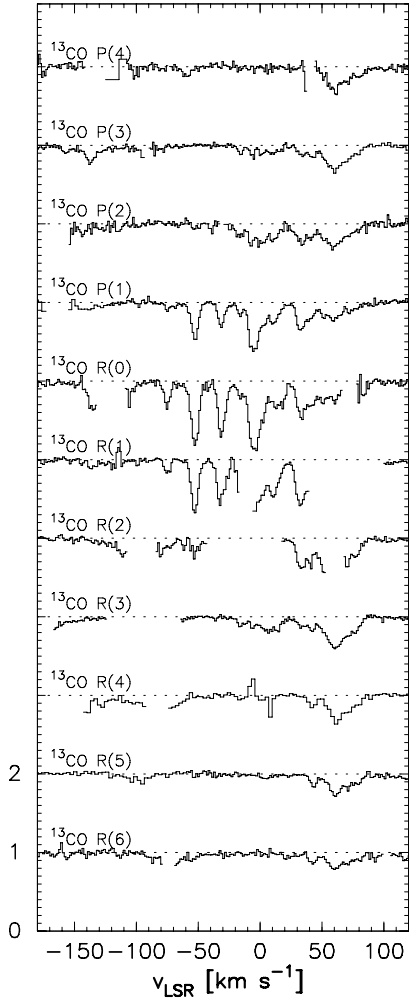}
\end{center}
  \caption{Spectra of individual $^{13}$CO~v=1-0 lines toward
    GCIRS~3.  Gaps in the individual profiles correspond to
    wavelengths intervals of poor atmospheric transmission
    and/or to wavelengths where $^{12}$CO v=1-0 lines overlap.}
  \label{fgr:population}
\end{figure}

Spectra toward GCIRS~3 and GCIRS~1W of the three H$_3^+$ lines
and the low-lying $^{13}$CO $P$(1) line are compared in
\ref{fgr:comparison}.  Spectra of a much wider range of
$^{13}$CO lines toward GCIRS~3 are shown in
\ref{fgr:population}. As can be seen in \ref{fgr:comparison},
with the exception of the density-sensitive H$_3^+$ $R$(2,2)$^l$
line, each line shows absorption over a wide velocity
range. From the foregoing discussion the presence of the
$R$(3,3)$^l$ line and absence of the $R$(2,2)$^l$ line over much
of this range implies that at most velocities where warm
molecular gas exists it is at low density.


For the most part the pairs of spectra of the other three lines
have nearly identical profiles over much of their velocity
ranges. This is not surprising, as both the Central Molecular
Zone and the intervening Galactic arms are enormous structures
compared to the separation of the GCIRS~3 and GCIRS~1W lines of
sight, $\sim$0.3~pc (one lightyear). The physical conditions in
the adjacent absorbing columns would not be expected to differ
much over such a small separation distance. On the other hand,
the Circumnuclear disk and gas physically associated with it are
much smaller so it would not be surprising to see differences in
that part of the absorption profile contributed by them.

Previous studies of H$_3^+$ and CO lines toward these and other
infrared sources in the Galactic center over much wider range of
sightlines than those shown here, together with knowledge of
Galactic structure gleaned mainly from radio observations, allow
one to clearly associate large portions of these line profiles
with gas in the Central Molecular Zone and the foreground arms
\cite{Oka:2005p8161, Goto:2008p981, Geballe:2010p10115}. The
narrow absorption features near 0, $-$30 and $-$50~km~s$^{-1}$,
common to the H$_3^+$ $R$(1,1)$^l$ and $^{13}$CO lines are
associated with cold and dense molecular gas in the foreground
Galactic arms. On the other hand, the shallow trough of
absorption seen in the $R$(1,1)$^l$ line profile to extend
roughly from $-$200 to 0~km~s$^{-1}$, nearly perfectly matched
by the full $R$(3,3)$^l$ profile, but conspicuously absent in
the both $R$(2,2)$^l$ and $^{13}$CO profiles, must be produced
by warm and low-density gas which, while containing H$_2$
(otherwise there would be no H$_3^+$), contains little or no
CO. This description is the calling card of diffuse molecular
gas, as introduced earlier in this paper. The wide spatial and
velocity extents of these absorptions and the warm temperature
of the gas require that the gas be located within the Central
Molecular Zone, and expanding outward from the center.

\begin{figure}[bth]
\begin{center}
\hspace*{-12mm}
\includegraphics[width=0.3\textwidth,angle=-90]{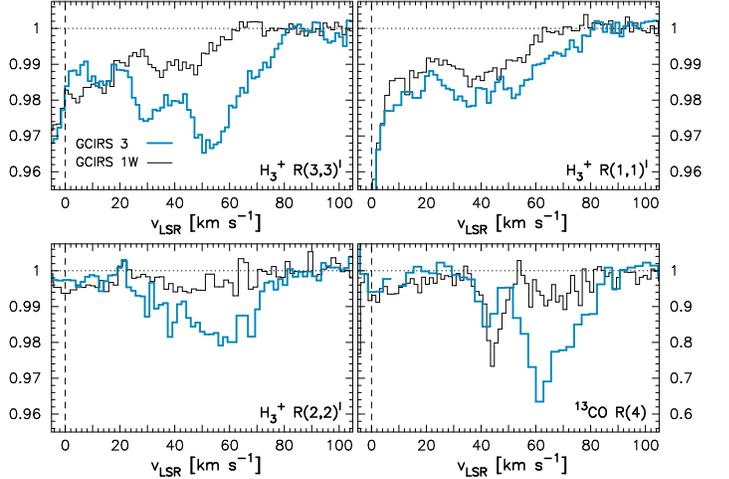}
\end{center}
\vspace{10mm}
 \caption{An expanded view of \ref{fgr:comparison} at the
   positive velocity 0--100~km\,s$^{-1}$. The velocity profile
   of the $^{13}$CO 1-0 $R$(4) line is shown, in place of the
   $P$(1) line.}
   \label{fgr:comparison-detail}
\end{figure}
\begin{figure}[bth]
\begin{center}
\hspace*{-15mm}
\includegraphics[width=0.4\textwidth,angle=0]{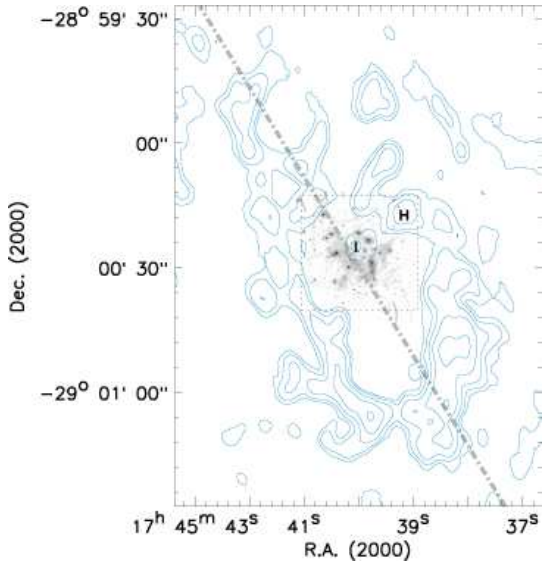}
\end{center}
  \caption{HCN~$J=$4-3 map of the Circumnuclear Disk
    \cite{MonteroCastano:2009p6848} (blue contours) overlaid
    with a VLT $K$-band image of the Central Cluster. The lowest
    contour level is 1.6~J\,m\,beam$^{-1}$\,s$^{-1}$
    (2\,$\sigma$). The dot-dash line denotes constant latitude
    and is parallel to the Galactic plane.  Clumps H and I, as
    identified by \citeauthor{MonteroCastano:2009p6848}, are
    labeled.}\label{fgr:hcn_map}
\end{figure}

\begin{figure}[bth]
\begin{center}
\includegraphics[width=0.5\textwidth,angle=0]{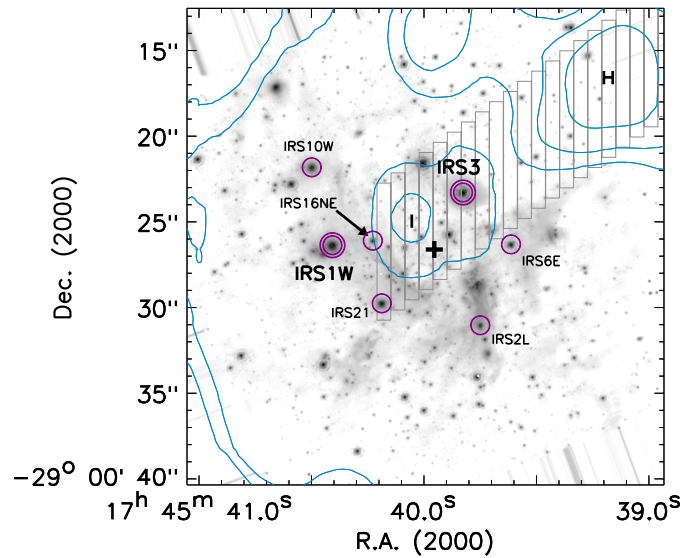}
\end{center}
  \caption{Expanded view of the central part of
    \ref{fgr:hcn_map}.  The cross is the position of Sgr A*. The
    positions of the apertures where the spectra in
    \ref{fgr:velocity} were extracted are marked with gray
    rectangles.}
  \label{fgr:blow-up}
\end{figure}


\subsection{Gas in the Central Cavity}


The largest differences between the line profiles toward
GCIRS~1W and GCIRS~3 in \ref{fgr:comparison} are at velocities
greater than about $+$20~km~s$^{-1}$. This velocity range is
shown in more detail in \ref{fgr:comparison-detail}; note that
in this figure we show the $^{13}$CO $R$(4) line as it is more
representative of the $^{13}$CO line profiles in this velocity
range than the $P$(1) line shown in \ref{fgr:comparison}.

In general the sightline toward GCIRS~3 produces stronger
absorptions than GCIRS~1W at $v~>$ $+$20 km~s$^{-1}$, but there
also are differences in the shapes of the profiles between the
two objects. Toward GCIRS~3 the absorption maxima of all the
three H$_3^+$ lines occur at $\sim+50$~km\,s$^{-1}$. Absorption
by all but the lowest lying $^{13}$CO lines also peak near
that velocity (see \ref{fgr:population} and the bottom right
panel of \ref{fgr:comparison-detail}).

Apart from GCIRS~1W and GCIRS~3 none of the observed sightlines
through the Central Molecular Zone within 30 pc of the center
produce absorption at $v >$$+$20~km~s$^{-1}$ in any of the
H$_3^+$ lines \cite{Goto:2008p981}.  Moreover, the presence of
both the excited H$_3^+$ lines and the $^{13}$CO absorption from
excited rotational levels (see \ref{fgr:population}) implies
that the absorbing gas is warm. Such absorptions cannot be
produced by the cold gas of the foreground spiral arms, but
instead must arise in the Galactic center. 

Despite the similarity in velocities we conclude that these
absorptions by H$_3^+$ and CO do not arise in the well-known
``$+$50~km\,s$^{-1}$'' giant molecular cloud
(M$-$0.02$-$0.07)\cite{Zylka:1990p6984,Mezger:1996p6971}, which
is located within the Central Molecular Zone, has dimensions of
roughly 20--30~pc, and is known from a multitude of radio
wavelength studies to extend across the sightlines to GCIRS~1W
and GCIRS~3.
The large differences between the absorption profiles at
positive velocities in GCIRS~3 and GCIRS~1W tends to rule out
the absorption occurring in such a large cloud.

This foregoing suggests that the absorption at positive
velocities occurs very close to GCIRS~3 and we thus consider the
possibility that the absorption arises in the Circumnuclear disk
and/or dense clouds that come off from the Circumnuclear disk to
the Central Cavity. \ref{fgr:hcn_map} is a contour map of the
Circumnuclear disk in the HCN~$J=$4-3 line obtained by
\citeauthor{MonteroCastano:2009p6848}\cite{MonteroCastano:2009p6848},
on which is superimposed a $K$-band ($\lambda=$2.2~$\mu$m) image
of the stars in the Central Cluster retrieved from the VLT
archive. As is shown in an expanded view (\ref{fgr:blow-up}) a
compact ``clump I'' (\citeauthor{MonteroCastano:2009p6848}),
which may be physically connected to the Circumnuclear disk,
positionally coincides with GCIRS~3, while the line of sight to
GCIRS~1W is clear. Note that, as $n_{\rm
  crit}\sim~10^{8}$~cm$^{-3}$ for the HCN~$J=$4-3 shown in
\ref{fgr:hcn_map}, this transition predominantly traces the
highest density portions of the disk. Lower density molecular
gas with much weaker HCN emission in this line could extend
considerably inward from the outermost contours.

\begin{figure}[bth]
\begin{center}
\vspace*{16mm}
\hspace*{-8mm}
\includegraphics[width=0.4\textwidth,angle=0]{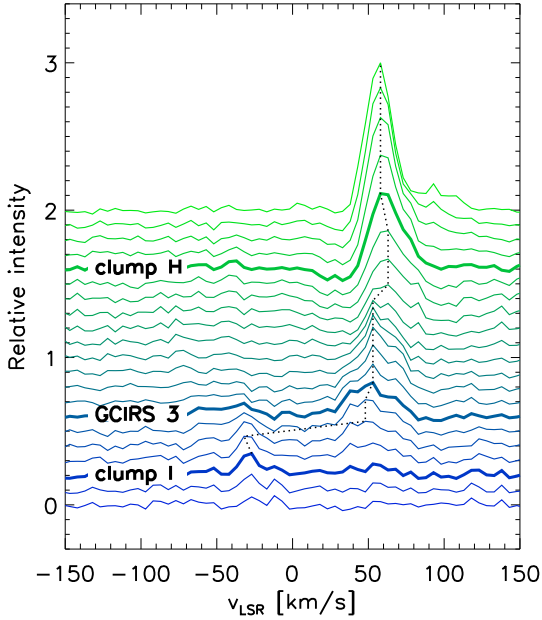}
\end{center}
  \caption{HCN~$J=$4-3 emission line spectra extracted from the
    apartures shown in \ref{fgr:blow-up} 
    from clump~H through the
    position of GCIRS~3 to clump~I. Velocities of maximum emission are
    connected by a dotted line.}
  \label{fgr:velocity}
\end{figure}

\begin{figure}[bth]
\begin{center}
\hspace*{-12mm}
\includegraphics[height=0.48\textwidth,angle=-90]{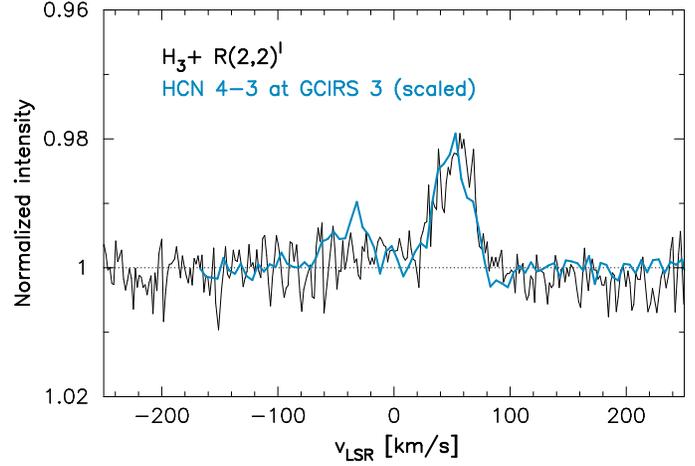}
\end{center}
\vspace{5mm}
  \caption{Comparison of the line profiles of H$_3^+$ $R$(2,2)$^l$ and 
     HCN~$J=$4-3 at the location of GCIRS~3. The HCN~$J=$4-3
    emission spectrum has been inverted and scaled to match the
    intensity of the $R$(2,2)$^l$ line.}
  \label{fgr:r2}
\end{figure}

To investigate the relationship between clump~I and the
Circumnuclear disk, HCN line emission spectra were extracted
along a line connecting the densest portion of the Circumnuclear
disk at clump~H \cite{MonteroCastano:2009p6848} with clump~I and
are shown in \ref{fgr:velocity}. From them it is clear that (1)
the Circumnuclear disk gas at clump H, seen at LSR velocities of
$+$50--$+$60~km~s$^{-1}$, extends inward across the sightline to
GCIRS~3 and close to clump I,  but not to GCIRS~1W, and (2)
clump I itself is mainly associated with gas at a radial
velocity of $\sim$~$-$30~km~s$^{-1}$. It would then appear that
$-$30~km~s$^{-1}$ gas in clump I is not physically connected to
the Disk.  As can be seen in \ref{fgr:r2} at GCIRS~3 the HCN
emission line profile near $+$50~km~s$^{-1}$ is an excellent
match to the H$_3^+$ $R$(2,2)$^l$ absorption line, not only in
its line center, but also in its line shape. We interpret this
as strong evidence that the absorption toward GCIRS~3 centered
at $+$50~km~s$^{-1}$ arises largely in this arm of molecular gas
extending inward to the west of clump~I from the Circumnuclear
disk.

\subsubsection{Temperature and density}

\begin{figure}[bth]
\begin{center}
\hspace*{-12mm}
\includegraphics[height=0.48\textwidth,angle=-90]{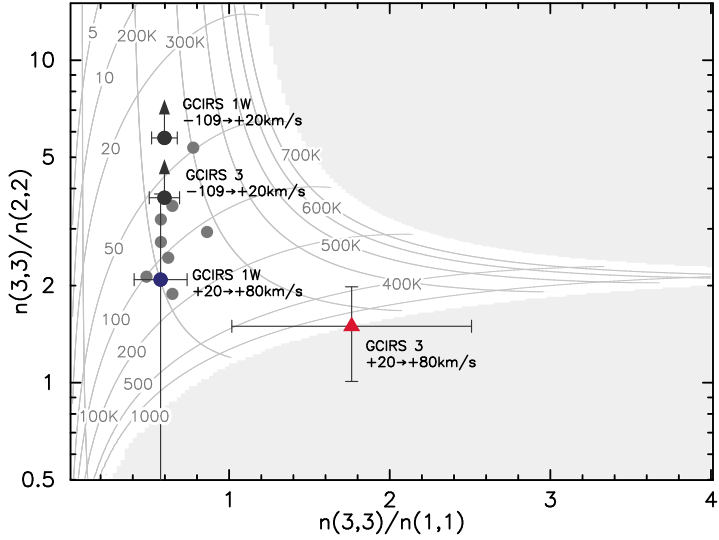}
\end{center}
  \caption{Comparison of the relative level populations of
    H$_3^+$, $n$(3,3)/$n$(2,2) and $n$(3,3)/$n$(1,1) in the lines
    of sight to GCIRS~3 and GCIRS~1W, derived from the steady state
    analysis by \citeauthor{Oka:2004p8755}\cite{Oka:2004p8755}.
    The gray dots correspond to other sightlines to sources in the
    Galactic center within 30~pc of Sgr~A* \cite{Goto:2008p981}
    that mostly sample the gas in the Central Molecular
    Zone. Note the higher temperatures and density in the gas in
    front of GCIRS~3 at radial velocities $+$20 to
    +80~km\,s$^{-1}$ (red triangle).}
\label{fgr:nt}
\end{figure}

We convert the measured equivalent widths of the H$_3^+$
absorption lines in the interval $+$20 to $+$80~km\,s$^{-1}$ to
column densities in the (1,1), (2,2) and (3,3) states using
standard procedures (see \citeauthor{Goto:2008p981}
\citeyear{Goto:2008p981}) and utilizing the dipole transition
moments given by
\citeauthor{Neale:1996p38895}\cite{Neale:1996p38895}.  The
  absorpiton arising in the diffuse clouds at the same velocity
  range, whose prsence is apparent in the spectra of GCIRS~1W,
  were removed by subtracting the spectra of GCIRS~1W from those
  of GCIRS~3 prior to the measuremment of the equivalent
  widths. Using the steady state analysis of
\citeauthor{Oka:2004p8755}\cite{Oka:2004p8755} one can then use
the relative level populations to estimate the temperature and
the density in the absorbing gas. We find 250~K$< T<$350~K and
$>$500~cm$^{-3}$ (\ref{fgr:nt}). This is sufficiently
dense for the gas to be classified as a dense cloud.
Unsurprisingly it is much warmer than the typical Galactic
equivalent.  The derived density is far less than the critical
density of the HCN $J=$4-3 transition, suggesting that observed
H$_3^+$ inhabits a different portion of the cloud than the
observed HCN, perhaps the outer layers. Detailed analysis of the
$^{13}$CO absorption feature centered at 60~km~s$^{-1}$ is
complicated, as the individual transitions have widely different
critical densities.  However, the relative line strengths
indicate temperatures exceeding 100~K and densities of $\sim
10^{6}$~cm$^{-3}$ (see \citeauthor{Kramer:2004p41581}
\cite{Kramer:2004p41581}
\citeyear{Kramer:2004p41581}).

\subsubsection{The cosmic ray ionization rate}

Here we derive the cosmic ray ionization rate in the
$+$50~km~s$^{-1}$ gas on the line of sight to GCIRS~3.   In
  Goto et al. 2008 $\zeta_2$ was calculated assuming the cloud
  at $+$50\,km\,s$^{-1}$ is diffuse, as in the case for the
  other clouds filling the large part of the Galactic center.
  The gas likely has the properties of a dense cloud (i.e, it is
  fully molecular), as is discussed above, and therefore we use
  equation (1), which can be re-expressed as

\[\zeta_2 L 
= 2k_L^\prime N({\rm H_3^+}) \frac{n_{\rm C}}{n_{\rm H}}
R_x,\]

\noindent where $L$ is the pathlength through the cloud, $N({\rm
  H_3^+})$ is the observed column density of H$_3^+$, $k_L^\prime = 
3\times 10^{-9}$~cm$^3$\,s$^{-1}$ is the Langevin rate constant for the 
reaction H$_3^+$ + CO $\rightarrow$ H$_2$ + HCO$^+$ taken from 
\citeauthor{Anicich:1986p38896}\cite{Anicich:1986p38896} and 
\citeauthor{Klippenstein:2010p41000}\cite{Klippenstein:2010p41000}, and 
multiplied by 1.5 to take into account the other minor destruction paths 
(especially H$_3^+$ + O $\rightarrow$ H$_2$ + OH$^+$ or $\rightarrow$ H 
+ H$_2$O$^+$), $\frac{n_{\rm
    C}}{n_{\rm H}}|_{SV} (= 1.6\times10^{-4})$
\cite{Sofia:2004p40486,Lacy:1994p24447} is the fractional
abundance of carbon relative to hydrogen in the solar vicinity,
and $R_x$ is the correction factor to take into account the
higher abundances of elements other than H in the Galactic
center.

We take $L$ to be 0.1~pc, roughly half the distance between
GCIRS~3, which lies behind the arm, and GCIRS~1W, where the
cloud is absent, and $R_x=3$, a conservatively low value in the
range proposed to date
\cite{Sodroski:1995p40498,Arimoto:1996p40506,Rolleston:2000p40514,Chiappini:2001p40528,Esteban:2005p40536}.
The total column density of H$_3^+$ is calculated by adding the
column density of $N$(1,1), $N$(2,2), and $N$(3,3), in addition
to $N$(1,0) published in 
  \citeauthor{Goto:2008p981}
  (\citeyear{Goto:2008p981})\cite{Goto:2008p981}. 
[$N({\rm H_3^+}) = 6 \times10^{14}$~cm$^{-2}$].  We thus obtain
$\zeta_2 = 1.2\times 10^{-15}$~s$^{-1}$.

We regard the above value for $\zeta_2$ as a lower limit,
because we have used a low value for $R_x$ and also have assumed
that the $+$50 km~s$^{-1}$ gas is fully molecular, which may
well be the case for the dense core which produces the HCN~
$J=$4-3 line emission, but may not be in the outer more exposed,
lower density regions where the bulk of the H$_3^+$ absorption
is likely to occur. The value of the ionization rate thus may
not differ greatly from values of
(2--7)~$\times~10^{-15}$~s$^{-1}$ derived for the Central
Molecular Zone as a whole by \citeauthor{Oka:2005p8161}
(\citeyear{Oka:2005p8161}).

\section{Sources of H$_2$ Ionization}

These values of the ionization rate are among the highest
measured in the Galaxy. It is therefore of interest to consider
possible sources of the ionizing particles. The black hole in
Sgr~A*, nearby supernova remnants, and energetic winds from
stars in the Central Cluster are all potential particle
accelerators. We have assumed so far that cosmic ray is the sole
source of ionization of H$_2$. This is probably the case for the
Central Molecular Zone, where the size of the medium
($\sim$200~pc) is much larger than the range of the X-ray
photons ($\sigma_x\approx 10^{-22}$\,cm$^{-2}$ at
  1~keV\cite{Wilms:2000p43635}), therefore X-ray does not
directly contribute to the ionization in much of the region, but
need not be the case in the central few parsecs. There the
ionization can be due to the local sources.  We discuss below
the H$_2$ ionization by the cosmic rays as well as by X-rays,
and estimate the expected $\zeta_2$ for each mechanism.

\subsection{Cosmic ray ionization}

The rate at which a single molecular hydrogen is ionized by an
energetic proton is given by the product of the proton 
  specific intensity, $J_p(E)$ and the ionization cross section
of H$_2$, $\sigma_p(E)$, integrated over energy and solid angle,

\[\zeta_2 = 4\,\pi\, \int_{E_1}^{E_2} J_p(E) \cdot \sigma_p(E) dE. \]

Ionization by the electron impact is neglected, since the energy
loss though radiative processes as a results of the interaction
with the interstellar magnetic field, ambient photons and
charged particles is much faster, and since the ionization cross
section is an order of magnitude smaller than for protons.

What kind of proton spectrum one should assume in the central
few parsecs of the Galaxy? A clue comes from a high energy (up
to $\sim$10~TeV) $\gamma$-ray observation. A point-like TeV
$\gamma$-ray source was recently discovered in the Galactic
center by the High Energy Stereoscopic System (HESS
Collaboration) \cite{Aharonian:2004p38239} which observes
optical Cherenkov light of an electron-positron pair generated
by the incidence of high energy $\gamma$-rays on the Earth's
atmosphere. The source, HESS~J1745$-$290, was subsequently
localized to within a few arcminutes of
Sgr~A*. \citeauthor{Aharonian:2005p34402}
\cite{Aharonian:2005p34402} argue that the most likely mechanism
to produce such high energy $\gamma$-rays is neutral pion decay,
precipitated by the acceleration of protons near the black
hole. When such protons encounter cold nuclei in the ambient
medium, proton-proton scattering produces neutral pions which
subsequently decays pairs of high energy $\gamma$-ray
photons. From the observed $\gamma$-ray spectrum, the energy
spectrum of protons injected to the cold medium can be
determined.
\citeauthor{Chernyakova:2011p34287}\cite{Chernyakova:2011p34287}
calculated that the proton flux required to reproduce the
observed $\gamma$-ray spectrum of HESS~J1745$-$290 from 0.1~GeV
to 100~TeV is $E^2F=1.4\times10^3$~ erg\,cm$^{-2}$\,s$^{-1}$ at
1~GeV in their representation, or 7$\times
10^4$~particles\,cm$^{-2}$\,s$^{-1}$\,str$^{-1}$\,(Gev/Nucleon)$^{-1}$.
For comparison, the standard proton flux in the Milky Way is
0.2~particles\,cm$^{-2}$\,s$^{-1}$\,str$^{-1}$\,(Gev/Nucleon)$^{-1}$
at 1~GeV \cite{Spitzer:1968p31337,Indriolo:2009p8835}, which is
$3\times 10^5$ times smaller.

\begin{figure}[bth]
\begin{center}
\hspace*{-12mm}
\includegraphics[height=0.48\textwidth,angle=-90]{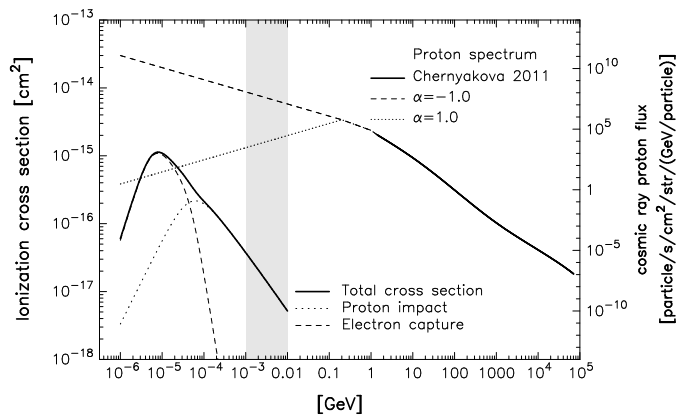}
\end{center}
\vspace{5mm}
  \caption{Proton energy distribution and ionization cross
    section of H$_2$ at 1~keV to 100~TeV. The spectrum is
    obtained from
    \citeauthor{Chernyakova:2011p34287}\cite{Chernyakova:2011p34287}
    at $E>$1~GeV, and is extrapolated in two ways from 0.2~GeV
    to 1 MeV, using $J_p(E)\propto E^{-1}$ and $J_p(E)\propto
    E$. The spectrum from 1~GeV to 0.2~GeV is bridged by
    $J_p(E)\propto E^{-1.25}$ to imitate a smooth break. See
    text for details.  The ionization cross section is
    reproduced from
    \citeauthor{Rudd:1983p36770}\cite{Rudd:1983p36770} and
    \citeauthor{Padovani:2009p31095}\cite{Padovani:2009p31095}.}
  \label{fgr:cosmic-ray}
\end{figure}

The collisional ionization of H$_2$ occurs via proton impact (H$_2$ + p 
$\rightarrow$ H$_2^+$ + p + e) and electron capture (H$_2$ + p 
$\rightarrow$ H$_2^+$ + H) and is well understood, theoretically 
\cite{Bethe:1933p39189} and experimentally \cite{Rudd:1983p36770}. The
combined ionization cross-section is reproduced from 
\citeauthor{Rudd:1983p36770,Padovani:2009p31095} 
\cite{Rudd:1983p36770,Padovani:2009p31095} in ~\ref{fgr:cosmic-ray}. As 
$\sigma_p(E)$ is well understood, the cosmic ray ionization rate is 
virtually only dependent on the proton spectrum $J_p(E)$. If $J_p(E)$ is 
scaled up by $3\times 10^5$ times from the Galactic standard, $\zeta_2$ 
becomes larger by the same factor. Such a large $\zeta_2$ has never been 
observed, not even in the Galactic center; the values that we and others 
have reported are only larger by factors of 10-100 than those outside 
the center.

There could be at least two possible explanations for this huge 
discrepancy: uncertain extrapolation of $J_p(E)$ to the lower energy, 
and the non-linear dependency of H$_3^+$ abundance on $\zeta_2$ for very 
high values of $\zeta_2$.

As mentioned earlier the cosmic ray spectrum cannot be directly observed 
in the energy range where ionization of H$_2$ is most efficient 
(1--10~MeV), because of the solar modulation. At much higher 
energies where direct observations are possible, the number flux of 
cosmic ray protons increases as energy decreases from TeV to GeV 
according to a power law, $J(E)\propto E^\alpha$, with $\alpha$ between 
$-$2 and $-$3. At lower energies the best measurements are by the 
Voyagers and  Pioneer spacecrafts in the outer solar system 
\cite{Webber:1998p31383,Webber:2009p39512}. Although the observations 
are still affected by the solar modulation, the local interstellar 
cosmic ray spectrum can be predicted using the latest solar modulation 
models \cite{Langner:2005p40106,Webber:2009p39512,Scherer:2011p40086}. 
The intrinsic cosmic ray spectra are roughly flat at 0.1--1~GeV
\cite{Moskalenko:2002p37919}, as is expected from the shorter range of 
the low energy cosmic rays ($2.5\times 10^{20}$~cm$^{-2}$ for 1~MeV 
cosmic ray proton \cite{Cravens:1978p38220}).

On the other hand, \citeauthor{Indriolo:2009p8835}
\cite{Indriolo:2009p8835} found that a low-energy turnover at
0.1--1~GeV is inconsistent with the mean ionization rate
inferred from H$_3^+$ spectroscopy. Instead the proton spectrum
must continue to increase down to 1~MeV, otherwise the H$_2$
ionization rate is too low compared to the observed
$\zeta_2=4\times 10^{-16}$~s$^{-1}$ in diffuse clouds, an
average value secured by H$_3^+$ spectroscopy toward more than a
dozen of sightlines \cite{Indriolo:2012p31279}. A similar
conclusion is reached by
\citeauthor{Neronov:2012p35437}\cite{Neronov:2012p35437}, who
analyzed $\gamma$-ray spectra of the molecular clouds close to
the solar system known as the Gould Belt clouds. The observed
$\gamma$-ray spectra were best reproduced by the cosmic-ray
protons having power-law spectrum with a weak break (a change of
the spectral index) at 9~GeV, but no turnover (change of sign of
the spectral index).

The uncertain extrapolation of the cosmic ray spectrum from the 
observable ($>$~1\,GeV) to unobservable energy ($\sim$MeV) is the common 
problem when we calculate $\zeta$ directly from $J_p(E)$. Guided by 
\citeauthor{Indriolo:2009p8835} and \citeauthor{Padovani:2009p31095}, we 
adapted two extreme power-law indices $\alpha=1$ and $-1$ at the low 
energy to extrapolate the proton spectrum from the break energy, which 
we set at 0.2~GeV. The former has a clear turnover, as indicated by the 
spacecraft measurements and the solar modulation models 
\cite{Moskalenko:2002p37919,Webber:2009p39512}. The latter contains a 
weak break but no turnover as is proposed by 
\citeauthor{Indriolo:2009p8835} and others 
\cite{Neronov:2012p35437,Nath:2012p34546}, Both cases are shown in 
\ref{fgr:cosmic-ray}. The proton injection spectrum by 
\citeauthor{Chernyakova:2011p34287} is well approximated by 
$J_p(E)\propto E^{-2.5}$ down to the energy 1~GeV. The gap 0.2--1~GeV 
was bridged by power-law spectrum with index $\alpha=-1.25$ to imitate a 
smooth break. The interval of the integration [$E_1$, $E_2$] was 
arbitrarily set to 1--10~MeV. The lower cut-off energy of the 
integration is often set at 1~MeV, because lower energy protons do not 
have sufficient range to affect on the global H$_2$ ionization rate. 
However, the ionization in this particular case can be local, and the 
cosmic rays with short range may well contribute to $\zeta_2$. The 
calculation therefore provides only a lower limit for $\zeta_2$.

 Integration yields $\zeta_2 = 5.5\times 10^{-11}$~s$^{-1}$
  for the non-turnover case with $\alpha=-1$ and a weak break
  and $1.6\times 10^{-14}$~s$^{-1}$ for $\alpha=1$ with a strong
  turnover. The numbers are reduced to $\zeta_2 = 4.6\times
  10^{-11}$~s$^{-1}$ and $\zeta_2 = 3.3\times
  10^{-15}$~s$^{-1}$, respectively, if the break energy is
  raised to 0.4~GeV. In either case the cosmic-ray proton
spectrum without a strong turnover is 4 orders of
magnitude higher than is observed in the cloud in front of
GCIRS~3.
This means that, if as many relativistic protons
are generated as the HESS TeV $\gamma$-ray source implies, they
must be strongly damped at low energy. The requirement for
damping is even stronger if the actual cut-off energy is lower
than 1~MeV.

A second possible explanation for the large discrepancy is that
the observed column density of H$_3^+$ might have led to a
significant underestimate of $\zeta_2$.  We have assumed that
the abundance of H$_3^+$ linearly increases with $\zeta_2$,
because the reaction of H$_2^+$ and H$_2$ is rapid enough to
justify neglecting other competing processes. At high $\zeta_2$
($>10^{-15}$~s$^{-1}$), however, destructive recombination of
H$_2^+$ with an electron becomes non-negligible, since electrons
are more populous because of the high $\zeta_2$
\cite{Liszt:2006p8116,Liszt:2007p33674}. It seems unlikely that
this can completely account for the discrepancy, however.

Finally, another factor that may need to be taken into account is the 
time-dependent formation rate of H$_2$, which would require 
$\sim$10$^7$~yrs to reach an equilibrium, if the formation starts from 
pure atomic gas \cite{Liszt:2007p33674}. The dynamical timescale of the 
Galactic center is short (the orbital period around the $4\times 10^6~ 
M_\odot$ black hole in Sgr~A* is 10,000~yrs at 1~pc) 
\cite{Genzel:2010p37134}. If the formation or destruction of H$_2$ is 
triggered by the changes of the external conditions of the cloud, the 
molecular hydrogen in the cloud in front of GCIRS~3 may not have enough 
time to reach steady state abundances balancing formation and 
destruction. At the beginning of the process, the abundance of H$_3^+$ 
is less dependent on $\zeta_2$, but more on the H$_2$ abundance. 
$\zeta_2$ calculated on the basis of the linear dependency on $N({\rm 
H_3^+})$ may not correctly represent the actual cosmic ray ionization 
rate.

\subsection{X-ray ionization}

\begin{figure}[bth]
\begin{center}
\hspace*{-12mm}
\includegraphics[height=0.48\textwidth,angle=-90]{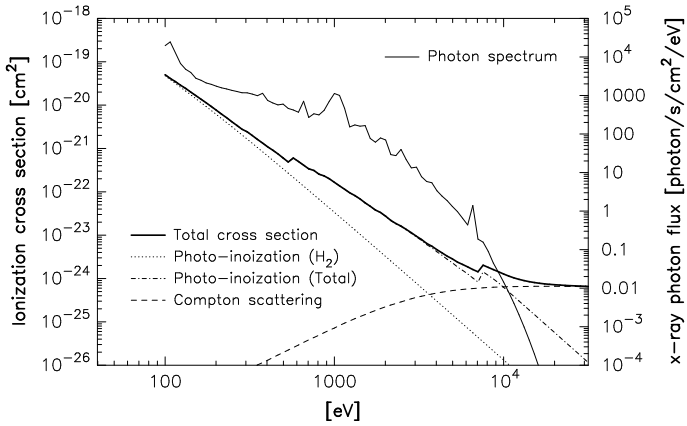}
\end{center}
\vspace{5mm}
  \caption{Calculated photon distribution of the diffuse local
    X-ray emission at a distance of 1~pc from Sgr~A*, and the
    ionization cross section of H$_2$, from 100~eV to
    10~keV. The spectrum was calculated from the Raymond-Smith
    plasma model \cite{Raymond:1977p38228} provided in 
      AtomDB\cite{Foster:2012p38116} with the plasma
    temperature $kT= 1.9$~keV, and scaled so that the
    luminosity matches to that of
    \citeauthor{Baganoff:2003p36758}\cite{Baganoff:2003p36758}
    modeled by
    \citeauthor{Rockefeller:2004p36525}\cite{Rockefeller:2004p36525}.
    The photo-ionization cross is taken from
    \citeauthor{Yan:1998p37828}\cite{Yan:1998p37828} and
    \citeauthor{Yan:2001p37834}\cite{Yan:2001p37834}, and
    Compton scattering cross section is taken from
    \citeauthor{Hubbell:1975p37862} \cite{Hubbell:1975p37862}.}
  \label{fgr:X-ray}
\end{figure}

Similar to the case of cosmic ray ionization, the X-ray
ionization rate is the integral over energy of the product of
the X-ray spectrum $F(E)$, and the ionization cross section
$\sigma_x(E)$. In the X-ray regime (0.1~keV$<E<$100~keV), two
ionization processes must be taken into account:
photo-ionization in which the X-ray photon is absorbed, and
Compton scattering in which the photon is scattered at lower
energy. In either case, the electron liberated by the ionization
is energetic enough to lead to further ionization events on
H$_2$. These secondary ionizations are far more numerous than
the first ionization by the X-ray photon. The
  photo-ionization cross section of H$_2$ by X-ray $\sigma_{xp}$
  is taken from \citeauthor{Yan:1998p37828}\cite{Yan:1998p37828}
  and \citeauthor{Yan:2001p37834}\cite{Yan:2001p37834}, and
  scaled by $\eta_{xp}$ according to
  \citeauthor{Wilms:2000p43635}\cite{Wilms:2000p43635} to take
  into account the contribution of the electrons produced by the
  photo-ionization of heavy elements. The Compton scattering
  cross section $\sigma_{xc}$ is taken from
  \citeauthor{Hubbell:1975p37862} \cite{Hubbell:1975p37862}, and
  scaled by $\eta_{xc}$ to include the electrons from the
  compton scattering on helium. The photo-ionization and compton
  scattering cross sections are reproduced in \ref{fgr:X-ray}.
The number of secondary ionizations that follow the single
primary ionization has been calculated by
\citeauthor{Dalgarno:1999p37811}\cite{Dalgarno:1999p37811} as a
mean energy per an ion pair $W$, where the total number of
ionization events per primary ionization $N_{\rm ion}$ is given
by $N_{\rm ion}=E_{\rm e}/W$, with $E_{\rm e}$ being the energy
of the initial (secondary) electron. Following
\citeauthor{Meijerink:2005p36531}\cite{Meijerink:2005p36531}, we
use $W$(1~keV) = 36\,eV over the entire energy range, as
$W$ asymptotically approaches this value for photons with E$>$1
keV \cite{Dalgarno:1999p37811}. The total X-ray ionization rate
is then given by

\[ \zeta_2 = \int{ (\sigma_{xp} \eta_{xp} + \sigma_{xc}\eta_{xc} ) \, \frac{E}{W({\rm 1 keV})}} F(E) \, dE.\]

Two possible sources of X-ray photons are present in the central
few parsecs of the Galaxy: X-ray point sources in the Central
Cavity and diffuse X-ray emission. At least five X-ray point
sources are known within 10 arcseconds of
Sgr~A*\cite{Baganoff:2003p36758}, including Sgr A* and GCIRS~13,
a compact star cluster that is thought to contain an
intermediate-mass black hole inside
\cite{Maillard:2004p38227}. All are of more or less similar
brightness ($10^{32}$--$10^{33}$~erg\,s$^{-1}$ at 2--10~keV)
with Sgr~A* slightly brighter than the others.

Here we calculate the ionization rate caused by X-ray
irradiation from Sgr~A* as an example. The actual $\zeta_2$
would be a few times larger if the contribution of other point
sources are included. The observed X-ray luminosity of Sgr~A* at
2--10~keV is $2.4\times 10^{33}$~erg\,s$^{-1}$
\cite{Baganoff:2003p36758}. In order to extrapolate the spectral
range to cover 0.1~keV to 100~keV, we create a spectrum using
the Raymond-Smith plasma model \cite{Raymond:1977p38228}
provided in AtomDB package \cite{Foster:2012p38116} with
the plasma temperature $kT=1.9$~keV as suggested by
\citeauthor{Baganoff:2003p36758}\cite{Baganoff:2003p36758} and
normalized to match the above observed luminosity. The complete
spectrum is shown in \ref{fgr:X-ray}. The flux of X-ray photons
that the cloud at GCIRS~3 receives is $F(E)= L_\ast(E)/4\pi
r^2$, where $r$ is the distance from Sgr~A* to the cloud,
assuming the X-ray emission is isotropic and that there is no
attenuation. The resultant $\zeta_2$ is highly dependent on the
distance to the X-ray source; using the minimum linear distance
between GCIRS~3 and Sgr~A* projected on the sky (=0.2~pc), 
  $\zeta_2$ is $5.8\times 10^{-13}$~s$^{-1}$ including the
  secondary ionizations, considerably larger than the value
  derived from the observed H$_3^+$ lines.

The second possible ionizing source is the local diffuse X-ray
emission that extends $\approx$10 arcseconds of Sgr~A*
\cite{Baganoff:2003p36758}. According to
\citeauthor{Rockefeller:2004p36525}\cite{Rockefeller:2004p36525}
the emission is likely the sum of the X-ray emission from the
stellar winds produced by the evolved massive stars in the
Central Cluster. Its X-ray luminosity is $7.6\times 10^{31}$~
erg\,s$^{-1}$\,arcsec$^{-2}$
(2--10~keV)\cite{Baganoff:2003p36758}. The plasma temperature
that best fits the observed spectrum is $kT=$1.3~keV. Note that
``the local diffuse X-ray emission'' is compact, and local to 10
arcseconds of Sgr~A*. It is distinguished from the large-scale
diffuse emission known as Galactic Ridge emission
\cite{Koyama:1996p39395}, and from the Sgr~A halo emission, or
from non-thermal emission from the supernova remnant Sgr~A~East,
which are not included in the above value.

To estimate the ionization rate from the diffuse X-ray emission
we assume a spherical X-ray emitting zone, centered on Sgr A*,
of radius 1~pc, somewhat less than the inner radius of the
Circumnuclear Disk, and locate the absorbing gas in front of
GCIRS~3 on the surface of this sphere.  The X-ray luminosity per
unit volume is modeled by
\citeauthor{Rockefeller:2004p36525}\cite{Rockefeller:2004p36525}
as a function of the distance from Sgr A*. We integrate the
luminosity per unit volume to determine the total luminosity
generated within the sphere.  Only the outward X-ray flux was
taken into account. The calculation yields $\zeta_2 = 4.8\times
10^{-13}$~s$^{-1}$ including the secondary ionizations, again
considerably larger than the value derived from the observed
H$_3^+$ lines.  We note that it is highly dependent on the
radius of the emitting sphere
(e.g., if the cloud is closer to the center (0.2~pc),
$\zeta_2$ is increased to $5.5\times 10^{-12}$~s$^{-1}$).

\subsection{Prospects and Status}

With sufficient numbers of suitable background sources such as
GCIRS~1W and GCIRS~3 in the central few parsecs of the Galaxy
and with a sufficiently extensive distribution of molecular gas
within the Central Cavity, one should be able to use
spectroscopy of H$_3^+$ to detect the footprints of the
ionization sources and to discriminate between the various
possibilities for ionization that have been put forth in the
previous section. If X-rays are the main source of the
ionization, $\zeta_2$ should be enhanced close to the individual
X-ray sources such as Sgr A* and the powerful wind source
GCIRS~13. If cosmic rays (protons) are the main ionization
source, $\zeta_2$ would be enhanced at the inner wall of the
Circumnuclear Disk, or on the compact clumps, where the
relativistic protons encounter cold nuclei. Inside the Central
Cavity, $\zeta_2$ would gradually decrease with the distance
from Sgr~A*, as the protons lose energy as they propagate
outward. High spatial sampling of $\zeta_2$ is the key to
address the primary ionization mechanism, and the possible in
situ acceleration of cosmic rays by the black hole.

\begin{figure}[bth]
\begin{center}
\hspace*{-30mm}
\includegraphics[height=0.75\textwidth,angle=-90]{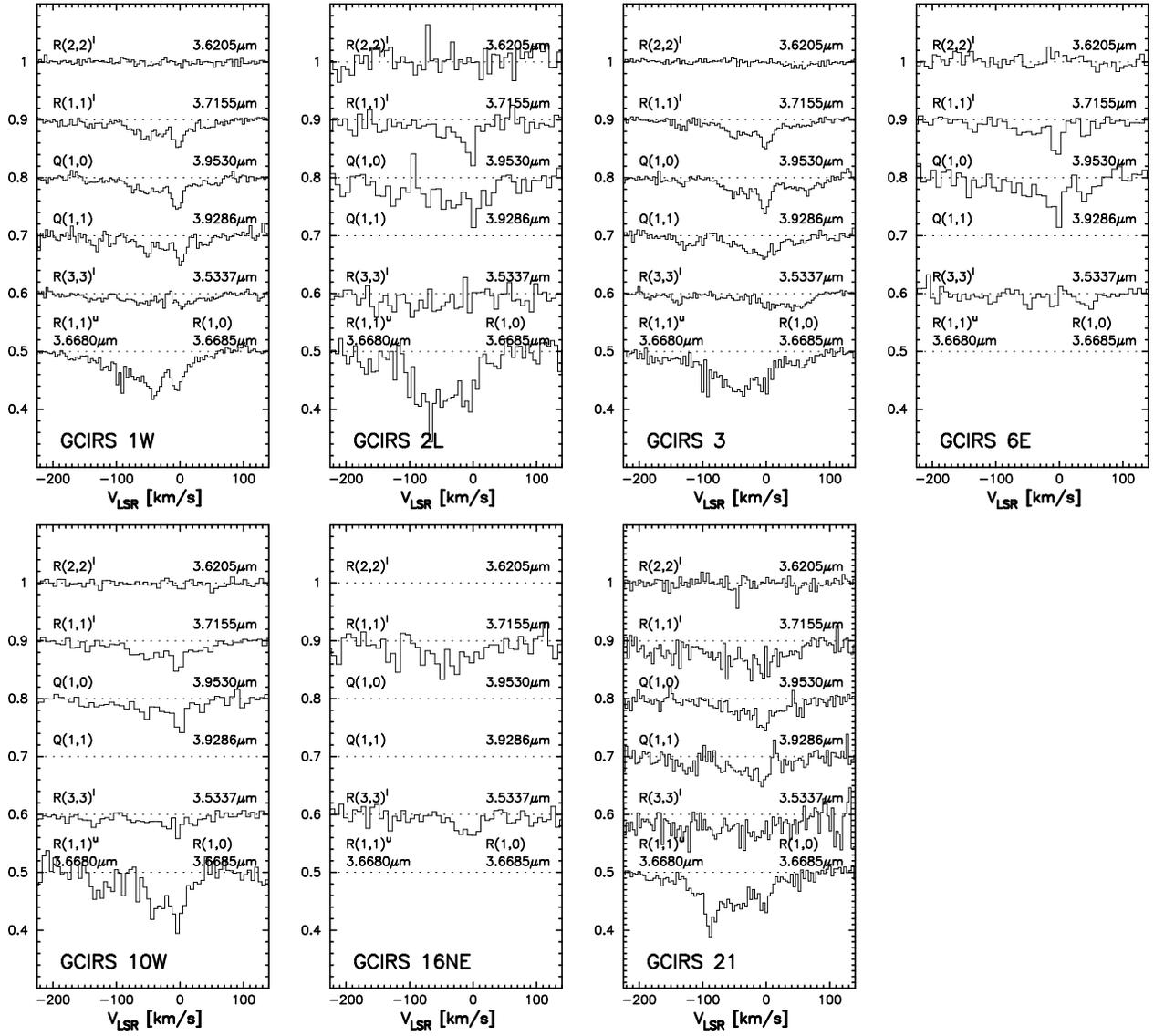}
\end{center}
\vspace{20mm}
\caption{H$_3^+$ spectra toward the stars in the Central Cluster
  obtained in the pilot survey carried out with IRCS at the
  Subaru Telescope.}
  \label{fgr:subaru}
\end{figure}

The Central Cluster contains hundreds of massive and luminous stars 
whose spectra are simple enough to allow quantitative spectroscopy of 
the H$_3^+$ lines from $J =$ 1, 2, and 3 levels that are needed to 
measure $\zeta_2$. The performances of current telescope and instruments 
limit the targets to a handful of stars brighter than $L<$7.5~mag (or 
$F_\lambda > 10^{-13}$~W\,m$^{-2}$\,$\mu$m$^{-1}$) (labeled in 
\ref{fgr:blow-up}). A pilot survey of H$_3^+$ toward these stars has 
been carried out at the Subaru Telescope on Mauna Kea with the 
spectrograph IRCS, and instrument that is efficient in surveying 
multiple lines. All sources so far studied show significant H$_3^+$ 
absorption from $J$=1 and 3 levels (\ref{fgr:subaru}). We intend to 
obtain improved spectra using CRIRES/VLT and to use it to search for the 
tracer of warm, dense clouds, the $R$(2,2)$^l$ line.

\onecolumn
\twocolumn

\begin{acknowledgement}

The authors thank the VLT and the Subaru Telescope for their
valuable assistance in obtaining the data. We thank Mar{\'i}a
ontero-Casta{\~n}o for providing HCN~$J=$4-3 map in the
machine-readable form. We thank Masha Chernyakova for the
ascertainment of the conversion of her proton energy spectrum to
the proton number count spectrum. We thank Andrii Neronov for
the explanation of his proton energy spectrum. The authors are
grateful to Takeshi Oka, not only for careful reading of an
early draft, but also for his continuous inspiration. M.G. is
supported by DFG grant GO 1927/3-1. T.R.G. is supported by the
Gemini Observatory, which is operated by the Association of
Universities for Research in Astronomy, Inc., on behalf of the
international Gemini partnership of Argentina, Australia,
Brazil, Canada, Chile, and the United States of America.


\end{acknowledgement}

\bibliography{cr}

\providecommand*\mcitethebibliography{\thebibliography}
\csname @ifundefined\endcsname{endmcitethebibliography}
  {\let\endmcitethebibliography\endthebibliography}{}
\begin{mcitethebibliography}{83}
\providecommand*\natexlab[1]{#1}
\providecommand*\mciteSetBstSublistMode[1]{}
\providecommand*\mciteSetBstMaxWidthForm[2]{}
\providecommand*\mciteBstWouldAddEndPuncttrue
  {\def\EndOfBibitem{\unskip.}}
\providecommand*\mciteBstWouldAddEndPunctfalse
  {\let\EndOfBibitem\relax}
\providecommand*\mciteSetBstMidEndSepPunct[3]{}
\providecommand*\mciteSetBstSublistLabelBeginEnd[3]{}
\providecommand*\EndOfBibitem{}
\mciteSetBstSublistMode{f}
\mciteSetBstMaxWidthForm{subitem}{(\alph{mcitesubitemcount})}
\mciteSetBstSublistLabelBeginEnd
  {\mcitemaxwidthsubitemform\space}
  {\relax}
  {\relax}

\bibitem[Herzberg(1969)]{Herzberg:1969p37807}
Herzberg,~G. Dissociation Energy and Ionization Potential of Molecular
  Hydrogen. \emph{Phys. Rev. Lett.} \textbf{1969}, \emph{23}, 1081--1083\relax
\mciteBstWouldAddEndPuncttrue
\mciteSetBstMidEndSepPunct{\mcitedefaultmidpunct}
{\mcitedefaultendpunct}{\mcitedefaultseppunct}\relax
\EndOfBibitem
\bibitem[Rudd et~al.(1983)Rudd, Goffe, Dubois, Toburen, and
  Ratcliffe]{Rudd:1983p36770}
Rudd,~M.~E.; Goffe,~T.~V.; Dubois,~R.~D.; Toburen,~L.~H.; Ratcliffe,~C.~A.
  Cross Sections for Ionization of Gases by 5 - 4000 keV Protons and for
  Electron Capture by 5 - 150 keV Protons. \emph{Phys. Rev. A: General Physics}
  \textbf{1983}, \emph{28}, 3244--3257\relax
\mciteBstWouldAddEndPuncttrue
\mciteSetBstMidEndSepPunct{\mcitedefaultmidpunct}
{\mcitedefaultendpunct}{\mcitedefaultseppunct}\relax
\EndOfBibitem
\bibitem[Padovani et~al.(2009)Padovani, Galli, and
  Glassgold]{Padovani:2009p31095}
Padovani,~M.; Galli,~D.; Glassgold,~A.~E. Cosmic-Ray Ionization of Molecular
  Clouds. \emph{Astron. Astrophys.} \textbf{2009}, \emph{501}, 619--631\relax
\mciteBstWouldAddEndPuncttrue
\mciteSetBstMidEndSepPunct{\mcitedefaultmidpunct}
{\mcitedefaultendpunct}{\mcitedefaultseppunct}\relax
\EndOfBibitem
\bibitem[Spitzer and Tomasko(1968)Spitzer, and Tomasko]{Spitzer:1968p31337}
Spitzer,~L.; Tomasko,~M.~G. Heating of H I Regions by Energetic Particles.
  \emph{Astrophys. J.} \textbf{1968}, \emph{152}, 971--986\relax
\mciteBstWouldAddEndPuncttrue
\mciteSetBstMidEndSepPunct{\mcitedefaultmidpunct}
{\mcitedefaultendpunct}{\mcitedefaultseppunct}\relax
\EndOfBibitem
\bibitem[Black and Dalgarno(1977)Black, and Dalgarno]{Black:1977p37990}
Black,~J.~H.; Dalgarno,~A. Models of Interstellar Clouds. I - The Zeta Ophiuchi
  Cloud. \emph{Astrophys. J., Suppl. Ser.} \textbf{1977}, \emph{34},
  405--423\relax
\mciteBstWouldAddEndPuncttrue
\mciteSetBstMidEndSepPunct{\mcitedefaultmidpunct}
{\mcitedefaultendpunct}{\mcitedefaultseppunct}\relax
\EndOfBibitem
\bibitem[van Dishoeck and Black(1986)van Dishoeck, and
  Black]{vanDishoeck:1986p37944}
van Dishoeck,~E.~F.; Black,~J.~H. Comprehensive Models of Diffuse Interstellar
  Clouds - Physical Conditions and Molecular Abundances. \emph{Astrophys. J.,
  Suppl. Ser.} \textbf{1986}, \emph{62}, 109--145\relax
\mciteBstWouldAddEndPuncttrue
\mciteSetBstMidEndSepPunct{\mcitedefaultmidpunct}
{\mcitedefaultendpunct}{\mcitedefaultseppunct}\relax
\EndOfBibitem
\bibitem[van~der Tak and van Dishoeck(2000)van~der Tak, and van
  Dishoeck]{vanderTak:2000p31226}
van~der Tak,~F. F.~S.; van Dishoeck,~E.~F. Limits on the Cosmic-Ray Ionization
  Rate toward Massive Young Stars. \emph{Astron. Astrophys.} \textbf{2000},
  \emph{358}, L79--L82\relax
\mciteBstWouldAddEndPuncttrue
\mciteSetBstMidEndSepPunct{\mcitedefaultmidpunct}
{\mcitedefaultendpunct}{\mcitedefaultseppunct}\relax
\EndOfBibitem
\bibitem[Geballe and Oka(1996)Geballe, and Oka]{Geballe:1996p8673}
Geballe,~T.~R.; Oka,~T. Detection of H$_3^+$ in Interstellar Space.
  \emph{Nature} \textbf{1996}, \emph{384}, 334--335\relax
\mciteBstWouldAddEndPuncttrue
\mciteSetBstMidEndSepPunct{\mcitedefaultmidpunct}
{\mcitedefaultendpunct}{\mcitedefaultseppunct}\relax
\EndOfBibitem
\bibitem[McCall et~al.(2002)McCall, Hinkle, Geballe, Moriarty-Schieven, Evans,
  Kawaguchi, Takano, Smith, and Oka]{McCall:2002p8292}
McCall,~B.~J.; Hinkle,~K.~H.; Geballe,~T.~R.; Moriarty-Schieven,~G.~H.;
  Evans,~N.~J.; Kawaguchi,~K.; Takano,~S.; Smith,~V.~V.; Oka,~T. Observations
  of H$_3^+$ in the Diffuse Interstellar Medium. \emph{Astrophys. J.}
  \textbf{2002}, \emph{567}, 391--406\relax
\mciteBstWouldAddEndPuncttrue
\mciteSetBstMidEndSepPunct{\mcitedefaultmidpunct}
{\mcitedefaultendpunct}{\mcitedefaultseppunct}\relax
\EndOfBibitem
\bibitem[Indriolo et~al.(2007)Indriolo, Geballe, Oka, and
  McCall]{Indriolo:2007p8061}
Indriolo,~N.; Geballe,~T.~R.; Oka,~T.; McCall,~B.~J. H$_3^+$ in Diffuse
  Interstellar Clouds: A Tracer for the Cosmic-Ray Ionization Rate.
  \emph{Astrophys. J.} \textbf{2007}, \emph{671}, 1736--1747\relax
\mciteBstWouldAddEndPuncttrue
\mciteSetBstMidEndSepPunct{\mcitedefaultmidpunct}
{\mcitedefaultendpunct}{\mcitedefaultseppunct}\relax
\EndOfBibitem
\bibitem[Indriolo and McCall(2012)Indriolo, and McCall]{Indriolo:2012p31279}
Indriolo,~N.; McCall,~B.~J. Investigating the Cosmic-Ray Ionization Rate in the
  Galactic Diffuse Interstellar Medium through Observations of H$_3^+$.
  \emph{Astrophys. J.} \textbf{2012}, \emph{745}, 91\relax
\mciteBstWouldAddEndPuncttrue
\mciteSetBstMidEndSepPunct{\mcitedefaultmidpunct}
{\mcitedefaultendpunct}{\mcitedefaultseppunct}\relax
\EndOfBibitem
\bibitem[Indriolo et~al.(2009)Indriolo, Fields, and McCall]{Indriolo:2009p8835}
Indriolo,~N.; Fields,~B.~D.; McCall,~B.~J. The Implications of a High
  Cosmic-Ray Ionization Rate in Diffuse Interstellar Clouds. \emph{Astrophys.
  J.} \textbf{2009}, \emph{694}, 257--267\relax
\mciteBstWouldAddEndPuncttrue
\mciteSetBstMidEndSepPunct{\mcitedefaultmidpunct}
{\mcitedefaultendpunct}{\mcitedefaultseppunct}\relax
\EndOfBibitem
\bibitem[Indriolo et~al.(2010)Indriolo, Blake, Goto, Usuda, Oka, Geballe,
  Fields, and McCall]{Indriolo:2010p31286}
Indriolo,~N.; Blake,~G.~A.; Goto,~M.; Usuda,~T.; Oka,~T.; Geballe,~T.~R.;
  Fields,~B.~D.; McCall,~B.~J. Investigating the Cosmic-Ray Ionization Rate
  Near the Supernova Remnant IC 443 through H$_3^+$ Observations.
  \emph{Astrophys. J.} \textbf{2010}, \emph{724}, 1357--1365\relax
\mciteBstWouldAddEndPuncttrue
\mciteSetBstMidEndSepPunct{\mcitedefaultmidpunct}
{\mcitedefaultendpunct}{\mcitedefaultseppunct}\relax
\EndOfBibitem
\bibitem[Yusef-Zadeh et~al.(2007)Yusef-Zadeh, Muno, Wardle, and
  Lis]{YusefZadeh:2007p7218}
Yusef-Zadeh,~F.; Muno,~M.; Wardle,~M.; Lis,~D.~C. The Origin of Diffuse X-Ray
  and $\gamma$-Ray Emission from the Galactic Center Region: Cosmic-Ray
  Particles. \emph{Astrophys. J.} \textbf{2007}, \emph{656}, 847--869\relax
\mciteBstWouldAddEndPuncttrue
\mciteSetBstMidEndSepPunct{\mcitedefaultmidpunct}
{\mcitedefaultendpunct}{\mcitedefaultseppunct}\relax
\EndOfBibitem
\bibitem[Tatischeff et~al.(2012)Tatischeff, Decourchelle, and
  Maurin]{Tatischeff:2012p41073}
Tatischeff,~V.; Decourchelle,~A.; Maurin,~G. Nonthermal X-Rays from Low-Energy
  Cosmic Rays: Application to the 6.4 keV Line Emission from the Arches Cluster
  Region. \emph{Astron. Astrophys.} \textbf{2012}, \emph{546}, 88\relax
\mciteBstWouldAddEndPuncttrue
\mciteSetBstMidEndSepPunct{\mcitedefaultmidpunct}
{\mcitedefaultendpunct}{\mcitedefaultseppunct}\relax
\EndOfBibitem
\bibitem[Becker et~al.(2011)Becker, Black, Safarzadeh, and
  Schuppan]{Becker:2011p31110}
Becker,~J.~K.; Black,~J.~H.; Safarzadeh,~M.; Schuppan,~F. Tracing the Sources
  of Cosmic Rays with Molecular Ions. \emph{Astrophys. J., Lett.}
  \textbf{2011}, \emph{739}, L43\relax
\mciteBstWouldAddEndPuncttrue
\mciteSetBstMidEndSepPunct{\mcitedefaultmidpunct}
{\mcitedefaultendpunct}{\mcitedefaultseppunct}\relax
\EndOfBibitem
\bibitem[Black(2012)]{Black:2012p37714}
Black,~J.~H. H$_3^+$ at the Interface between Astrochemistry and Astroparticle
  Physics. \emph{Philos. Trans. R. Soc., A} \textbf{2012}, \emph{370},
  5130--5141\relax
\mciteBstWouldAddEndPuncttrue
\mciteSetBstMidEndSepPunct{\mcitedefaultmidpunct}
{\mcitedefaultendpunct}{\mcitedefaultseppunct}\relax
\EndOfBibitem
\bibitem[Goto et~al.(2008)Goto, Usuda, Nagata, Geballe, McCall, Indriolo, Suto,
  Henning, Morong, and Oka]{Goto:2008p981}
Goto,~M.; Usuda,~T.; Nagata,~T.; Geballe,~T.~R.; McCall,~B.~J.; Indriolo,~N.;
  Suto,~H.; Henning,~T.; Morong,~C.~P.; Oka,~T. Absorption Line Survey of
  H$_3^+$ toward the Galactic Center Sources. II. Eight Infrared Sources within
  30 pc of the Galactic Center. \emph{Astrophys. J.} \textbf{2008}, \emph{688},
  306--319\relax
\mciteBstWouldAddEndPuncttrue
\mciteSetBstMidEndSepPunct{\mcitedefaultmidpunct}
{\mcitedefaultendpunct}{\mcitedefaultseppunct}\relax
\EndOfBibitem
\bibitem[Crocker et~al.(2011)Crocker, Jones, Aharonian, Law, Melia, Oka, and
  Ott]{Crocker:2011p35207}
Crocker,~R.~M.; Jones,~D.~I.; Aharonian,~F.; Law,~C.~J.; Melia,~F.; Oka,~T.;
  Ott,~J. Wild at Heart: the Particle Astrophysics of the Galactic Centre.
  \emph{Mon. Not. R. Astron. Soc.} \textbf{2011}, \emph{413}, 763--788\relax
\mciteBstWouldAddEndPuncttrue
\mciteSetBstMidEndSepPunct{\mcitedefaultmidpunct}
{\mcitedefaultendpunct}{\mcitedefaultseppunct}\relax
\EndOfBibitem
\bibitem[Tammann et~al.(1994)Tammann, Loeffler, and
  Schroeder]{Tammann:1994p40281}
Tammann,~G.~A.; Loeffler,~W.; Schroeder,~A. The Galactic Supernova Rate.
  \emph{Astrophys. J., Suppl. Ser.} \textbf{1994}, \emph{92}, 487--493\relax
\mciteBstWouldAddEndPuncttrue
\mciteSetBstMidEndSepPunct{\mcitedefaultmidpunct}
{\mcitedefaultendpunct}{\mcitedefaultseppunct}\relax
\EndOfBibitem
\bibitem[van~den Bergh and Tammann(1991)van~den Bergh, and
  Tammann]{vandenBergh:1991p40259}
van~den Bergh,~S.; Tammann,~G.~A. Galactic and Extragalactic Supernova Rates.
  \emph{Annu. Rev. Astron. Astrophys.} \textbf{1991}, \emph{29}, 363--407\relax
\mciteBstWouldAddEndPuncttrue
\mciteSetBstMidEndSepPunct{\mcitedefaultmidpunct}
{\mcitedefaultendpunct}{\mcitedefaultseppunct}\relax
\EndOfBibitem
\bibitem[Morris and Serabyn(1996)Morris, and Serabyn]{Morris:1996p19505}
Morris,~M.; Serabyn,~E. The Galactic Center Environment. \emph{Annu. Rev.
  Astron. Astrophys.} \textbf{1996}, \emph{34}, 645--701\relax
\mciteBstWouldAddEndPuncttrue
\mciteSetBstMidEndSepPunct{\mcitedefaultmidpunct}
{\mcitedefaultendpunct}{\mcitedefaultseppunct}\relax
\EndOfBibitem
\bibitem[Goto et~al.(2002)Goto, McCall, Geballe, Usuda, Kobayashi, Terada, and
  Oka]{Goto:2002p1051}
Goto,~M.; McCall,~B.~J.; Geballe,~T.~R.; Usuda,~T.; Kobayashi,~N.; Terada,~H.;
  Oka,~T. Absorption Line Survey of H$_3^+$ toward the Galactic Center Sources
  I. GCS 3-2 and GC IRS3. \emph{Publ. Astron. Soc. Jpn.} \textbf{2002},
  \emph{54}, 951--961\relax
\mciteBstWouldAddEndPuncttrue
\mciteSetBstMidEndSepPunct{\mcitedefaultmidpunct}
{\mcitedefaultendpunct}{\mcitedefaultseppunct}\relax
\EndOfBibitem
\bibitem[Oka et~al.(2005)Oka, Geballe, Goto, Usuda, and McCall]{Oka:2005p8161}
Oka,~T.; Geballe,~T.~R.; Goto,~M.; Usuda,~T.; McCall,~B.~J. Hot and Diffuse
  Clouds near the Galactic Center Probed by Metastable H$_3^+$.
  \emph{Astrophys. J.} \textbf{2005}, \emph{632}, 882--893\relax
\mciteBstWouldAddEndPuncttrue
\mciteSetBstMidEndSepPunct{\mcitedefaultmidpunct}
{\mcitedefaultendpunct}{\mcitedefaultseppunct}\relax
\EndOfBibitem
\bibitem[Gillessen et~al.(2009)Gillessen, Eisenhauer, Trippe, Alexander,
  Genzel, Martins, and Ott]{Gillessen:2009p6070}
Gillessen,~S.; Eisenhauer,~F.; Trippe,~S.; Alexander,~T.; Genzel,~R.;
  Martins,~F.; Ott,~T. Monitoring Stellar Orbits Around the Massive Black Hole
  in the Galactic Center. \emph{Astrophys. J.} \textbf{2009}, \emph{692},
  1075--1109\relax
\mciteBstWouldAddEndPuncttrue
\mciteSetBstMidEndSepPunct{\mcitedefaultmidpunct}
{\mcitedefaultendpunct}{\mcitedefaultseppunct}\relax
\EndOfBibitem
\bibitem[Krabbe et~al.(1995)Krabbe, Genzel, Eckart, Najarro, Lutz, Cameron,
  Kroker, Tacconi-Garman, Thatte, Weitzel, and et~al.]{Krabbe:1995p7344}
Krabbe,~A.; Genzel,~R.; Eckart,~A.; Najarro,~F.; Lutz,~D.; Cameron,~M.;
  Kroker,~H.; Tacconi-Garman,~L.~E.; Thatte,~N.; Weitzel,~L.; et~al., The
  Nuclear Cluster of the Milky Way: Star Formation and Velocity Dispersion in
  the Central 0.5 Parsec. \emph{Astrophys. J., Lett.} \textbf{1995},
  \emph{447}, L95--L99\relax
\mciteBstWouldAddEndPuncttrue
\mciteSetBstMidEndSepPunct{\mcitedefaultmidpunct}
{\mcitedefaultendpunct}{\mcitedefaultseppunct}\relax
\EndOfBibitem
\bibitem[Christopher et~al.(2005)Christopher, Scoville, Stolovy, and
  Yun]{Christopher:2005p7283}
Christopher,~M.~H.; Scoville,~N.~Z.; Stolovy,~S.~R.; Yun,~M.~S. HCN and HCO$^+$
  Observations of the Galactic Circumnuclear Disk. \emph{Astrophys. J.}
  \textbf{2005}, \emph{622}, 346--365\relax
\mciteBstWouldAddEndPuncttrue
\mciteSetBstMidEndSepPunct{\mcitedefaultmidpunct}
{\mcitedefaultendpunct}{\mcitedefaultseppunct}\relax
\EndOfBibitem
\bibitem[Montero-Casta{\~n}o et~al.(2009)Montero-Casta{\~n}o, Herrnstein, and
  Ho]{MonteroCastano:2009p6848}
Montero-Casta{\~n}o,~M.; Herrnstein,~R.~M.; Ho,~P. T.~P. Gas Infall Toward Sgr
  A* from the Clumpy Circumnuclear Disk. \emph{Astrophys. J.} \textbf{2009},
  \emph{695}, 1477--1494\relax
\mciteBstWouldAddEndPuncttrue
\mciteSetBstMidEndSepPunct{\mcitedefaultmidpunct}
{\mcitedefaultendpunct}{\mcitedefaultseppunct}\relax
\EndOfBibitem
\bibitem[G{\"u}sten(1987)]{Gusten:1987p19488}
G{\"u}sten,~R. Atomic and Molecular Gas in the Circumnuclear Disk. \emph{AIP
  Conf. Proc.} \textbf{1987}, \emph{155}, 19\relax
\mciteBstWouldAddEndPuncttrue
\mciteSetBstMidEndSepPunct{\mcitedefaultmidpunct}
{\mcitedefaultendpunct}{\mcitedefaultseppunct}\relax
\EndOfBibitem
\bibitem[Requena-Torres et~al.(2012)Requena-Torres, G{\"u}sten, Wei{\ss},
  Harris, Mart{\'\i}n-Pintado, Stutzki, Klein, Heyminck, and
  Risacher]{RequenaTorres:2012p40543}
Requena-Torres,~M.~A.; G{\"u}sten,~R.; Wei{\ss},~A.; Harris,~A.~I.;
  Mart{\'\i}n-Pintado,~J.; Stutzki,~J.; Klein,~B.; Heyminck,~S.; Risacher,~C.
  GREAT Confirms Transient Nature of the Circum-nuclear Disk. \emph{Astron.
  Astrophys.} \textbf{2012}, \emph{542}, L21\relax
\mciteBstWouldAddEndPuncttrue
\mciteSetBstMidEndSepPunct{\mcitedefaultmidpunct}
{\mcitedefaultendpunct}{\mcitedefaultseppunct}\relax
\EndOfBibitem
\bibitem[Baganoff et~al.(2003)Baganoff, Maeda, Morris, Bautz, Brandt, Cui,
  Doty, Feigelson, Garmire, Pravdo, and et~al.]{Baganoff:2003p36758}
Baganoff,~F.~K.; Maeda,~Y.; Morris,~M.; Bautz,~M.~W.; Brandt,~W.~N.; Cui,~W.;
  Doty,~J.~P.; Feigelson,~E.~D.; Garmire,~G.~P.; Pravdo,~S.~H.; et~al., Chandra
  X-Ray Spectroscopic Imaging of Sagittarius A* and the Central Parsec of the
  Galaxy. \emph{Astrophys. J.} \textbf{2003}, \emph{591}, 891--915\relax
\mciteBstWouldAddEndPuncttrue
\mciteSetBstMidEndSepPunct{\mcitedefaultmidpunct}
{\mcitedefaultendpunct}{\mcitedefaultseppunct}\relax
\EndOfBibitem
\bibitem[Becklin and Neugebauer(1968)Becklin, and
  Neugebauer]{Becklin:1968p40541}
Becklin,~E.~E.; Neugebauer,~G. Infrared Observations of the Galactic Center.
  \emph{Astrophys. J.} \textbf{1968}, \emph{151}, 145--161\relax
\mciteBstWouldAddEndPuncttrue
\mciteSetBstMidEndSepPunct{\mcitedefaultmidpunct}
{\mcitedefaultendpunct}{\mcitedefaultseppunct}\relax
\EndOfBibitem
\bibitem[Viehmann et~al.(2005)Viehmann, Eckart, Sch{\"o}del, Moultaka,
  Straubmeier, and Pott]{Viehmann:2005p5622}
Viehmann,~T.; Eckart,~A.; Sch{\"o}del,~R.; Moultaka,~J.; Straubmeier,~C.;
  Pott,~J.-U. L- and M-Band Imaging Observations of the Galactic Center Region.
  \emph{Astron. Astrophys.} \textbf{2005}, \emph{433}, 117--125\relax
\mciteBstWouldAddEndPuncttrue
\mciteSetBstMidEndSepPunct{\mcitedefaultmidpunct}
{\mcitedefaultendpunct}{\mcitedefaultseppunct}\relax
\EndOfBibitem
\bibitem[Paumard et~al.(2006)Paumard, Genzel, Martins, Nayakshin, Beloborodov,
  Levin, Trippe, Eisenhauer, Ott, Gillessen, and et~al.]{Paumard:2006p5913}
Paumard,~T.; Genzel,~R.; Martins,~F.; Nayakshin,~S.; Beloborodov,~A.~M.;
  Levin,~Y.; Trippe,~S.; Eisenhauer,~F.; Ott,~T.; Gillessen,~S.; et~al., The
  Two Young Star Disks in the Central Parsec of the Galaxy: Properties,
  Dynamics, and Formation. \emph{Astrophys. J.} \textbf{2006}, \emph{643},
  1011--1035\relax
\mciteBstWouldAddEndPuncttrue
\mciteSetBstMidEndSepPunct{\mcitedefaultmidpunct}
{\mcitedefaultendpunct}{\mcitedefaultseppunct}\relax
\EndOfBibitem
\bibitem[Najarro et~al.(1997)Najarro, Krabbe, Genzel, Lutz, Kudritzki, and
  Hillier]{Najarro:1997p6043}
Najarro,~F.; Krabbe,~A.; Genzel,~R.; Lutz,~D.; Kudritzki,~R.~P.; Hillier,~D.~J.
  Quantitative Spectroscopy of the HeI Cluster in the Galactic Center.
  \emph{Astron. Astrophys.} \textbf{1997}, \emph{325}, 700--708\relax
\mciteBstWouldAddEndPuncttrue
\mciteSetBstMidEndSepPunct{\mcitedefaultmidpunct}
{\mcitedefaultendpunct}{\mcitedefaultseppunct}\relax
\EndOfBibitem
\bibitem[Martins et~al.(2007)Martins, Genzel, Hillier, Eisenhauer, Paumard,
  Gillessen, Ott, and Trippe]{Martins:2007p7448}
Martins,~F.; Genzel,~R.; Hillier,~D.~J.; Eisenhauer,~F.; Paumard,~T.;
  Gillessen,~S.; Ott,~T.; Trippe,~S. Stellar and Wind Properties of Massive
  Stars in the Central Parsec of the Galaxy. \emph{Astron. Astrophys.}
  \textbf{2007}, \emph{468}, 233--254\relax
\mciteBstWouldAddEndPuncttrue
\mciteSetBstMidEndSepPunct{\mcitedefaultmidpunct}
{\mcitedefaultendpunct}{\mcitedefaultseppunct}\relax
\EndOfBibitem
\bibitem[Zhao et~al.(2009)Zhao, Morris, Goss, and An]{Zhao:2009p5799}
Zhao,~J.-H.; Morris,~M.~R.; Goss,~W.~M.; An,~T. Dynamics of Ionized Gas at the
  Galactic Center: Very Large Array Observations of the Three-Dimensional
  Velocity Field and Location of the Ionized Streams in Sagittarius A West.
  \emph{Astrophys. J.} \textbf{2009}, \emph{699}, 186\relax
\mciteBstWouldAddEndPuncttrue
\mciteSetBstMidEndSepPunct{\mcitedefaultmidpunct}
{\mcitedefaultendpunct}{\mcitedefaultseppunct}\relax
\EndOfBibitem
\bibitem[Zhao et~al.(2010)Zhao, Blundell, Moran, Downes, Schuster, and
  Marrone]{Zhao:2010p40563}
Zhao,~J.-H.; Blundell,~R.; Moran,~J.~M.; Downes,~D.; Schuster,~K.~F.;
  Marrone,~D.~P. The High-density Ionized Gas in the Central Parsec of the
  Galaxy. \emph{Astrophys. J.} \textbf{2010}, \emph{723}, 1097--1109\relax
\mciteBstWouldAddEndPuncttrue
\mciteSetBstMidEndSepPunct{\mcitedefaultmidpunct}
{\mcitedefaultendpunct}{\mcitedefaultseppunct}\relax
\EndOfBibitem
\bibitem[Kaeufl et~al.(2004)Kaeufl, Ballester, Biereichel, Delabre, Donaldson,
  Dorn, Fedrigo, Finger, Fischer, Franza, and et~al.]{Kaeufl:2004p40429}
Kaeufl,~H.-U.; Ballester,~P.; Biereichel,~P.; Delabre,~B.; Donaldson,~R.;
  Dorn,~R.; Fedrigo,~E.; Finger,~G.; Fischer,~G.; Franza,~F.; et~al., CRIRES: a
  High-Resolution Infrared Spectrograph for ESO's VLT. \emph{Ground-based
  Instrumentation for Astronomy. Edited by Alan F. M. Moorwood and Iye
  Masanori. Proceedings of the SPIE} \textbf{2004}, \emph{5492},
  1218--1227\relax
\mciteBstWouldAddEndPuncttrue
\mciteSetBstMidEndSepPunct{\mcitedefaultmidpunct}
{\mcitedefaultendpunct}{\mcitedefaultseppunct}\relax
\EndOfBibitem
\bibitem[Eisenhauer et~al.(2005)Eisenhauer, Genzel, Alexander, Abuter, Paumard,
  Ott, Gilbert, Gillessen, Horrobin, Trippe, and et~al.]{Eisenhauer:2005p6146}
Eisenhauer,~F.; Genzel,~R.; Alexander,~T.; Abuter,~R.; Paumard,~T.; Ott,~T.;
  Gilbert,~A.; Gillessen,~S.; Horrobin,~M.; Trippe,~S.; et~al., SINFONI in the
  Galactic Center: Young Stars and Infrared Flares in the Central Light-Month.
  \emph{Astrophys. J.} \textbf{2005}, \emph{628}, 246--259\relax
\mciteBstWouldAddEndPuncttrue
\mciteSetBstMidEndSepPunct{\mcitedefaultmidpunct}
{\mcitedefaultendpunct}{\mcitedefaultseppunct}\relax
\EndOfBibitem
\bibitem[Ghez et~al.(2008)Ghez, Salim, Weinberg, Lu, Do, Dunn, Matthews,
  Morris, Yelda, Becklin, and et~al.]{Ghez:2008p41514}
Ghez,~A.~M.; Salim,~S.; Weinberg,~N.~N.; Lu,~J.~R.; Do,~T.; Dunn,~J.~K.;
  Matthews,~K.; Morris,~M.~R.; Yelda,~S.; Becklin,~E.~E.; et~al., Measuring
  Distance and Properties of the Milky Way's Central Supermassive Black Hole
  with Stellar Orbits. \emph{Astrophys. J.} \textbf{2008}, \emph{689},
  1044--1062\relax
\mciteBstWouldAddEndPuncttrue
\mciteSetBstMidEndSepPunct{\mcitedefaultmidpunct}
{\mcitedefaultendpunct}{\mcitedefaultseppunct}\relax
\EndOfBibitem
\bibitem[Geballe and Oka(2010)Geballe, and Oka]{Geballe:2010p10115}
Geballe,~T.~R.; Oka,~T. Two New and Remarkable Sightlines Through the Galactic
  Center's Molecular Gas. \emph{Astrophys. J., Lett.} \textbf{2010},
  \emph{709}, L70--L73\relax
\mciteBstWouldAddEndPuncttrue
\mciteSetBstMidEndSepPunct{\mcitedefaultmidpunct}
{\mcitedefaultendpunct}{\mcitedefaultseppunct}\relax
\EndOfBibitem
\bibitem[Zylka et~al.(1990)Zylka, Mezger, and Wink]{Zylka:1990p6984}
Zylka,~R.; Mezger,~P.~G.; Wink,~J.~E. Anatomy of the Sagittarius A Complex. I -
  Geometry, Morphology and Dynamics of the Central 50 to 100 pc. \emph{Astron.
  Astrophys.} \textbf{1990}, \emph{234}, 133--146\relax
\mciteBstWouldAddEndPuncttrue
\mciteSetBstMidEndSepPunct{\mcitedefaultmidpunct}
{\mcitedefaultendpunct}{\mcitedefaultseppunct}\relax
\EndOfBibitem
\bibitem[Mezger et~al.(1996)Mezger, Duschl, and Zylka]{Mezger:1996p6971}
Mezger,~P.~G.; Duschl,~W.~J.; Zylka,~R. The Galactic Center: a Laboratory for
  AGN? \emph{The Astronomy and Astrophysics Review} \textbf{1996}, \emph{7},
  289--388\relax
\mciteBstWouldAddEndPuncttrue
\mciteSetBstMidEndSepPunct{\mcitedefaultmidpunct}
{\mcitedefaultendpunct}{\mcitedefaultseppunct}\relax
\EndOfBibitem
\bibitem[Neale et~al.(1996)Neale, Miller, and Tennyson]{Neale:1996p38895}
Neale,~L.; Miller,~S.; Tennyson,~J. Spectroscopic Properties of the H$_3^+$
  Molecule: A New Calculated Line List. \emph{Astrophys. J.} \textbf{1996},
  \emph{464}, 516--520\relax
\mciteBstWouldAddEndPuncttrue
\mciteSetBstMidEndSepPunct{\mcitedefaultmidpunct}
{\mcitedefaultendpunct}{\mcitedefaultseppunct}\relax
\EndOfBibitem
\bibitem[Oka and Epp(2004)Oka, and Epp]{Oka:2004p8755}
Oka,~T.; Epp,~E. The Nonthermal Rotational Distribution of H$_3^+$.
  \emph{Astrophys. J.} \textbf{2004}, \emph{613}, 349--354\relax
\mciteBstWouldAddEndPuncttrue
\mciteSetBstMidEndSepPunct{\mcitedefaultmidpunct}
{\mcitedefaultendpunct}{\mcitedefaultseppunct}\relax
\EndOfBibitem
\bibitem[Kramer et~al.(2004)Kramer, Jakob, Mookerjea, Schneider, Br{\"u}ll, and
  Stutzki]{Kramer:2004p41581}
Kramer,~C.; Jakob,~H.; Mookerjea,~B.; Schneider,~N.; Br{\"u}ll,~M.; Stutzki,~J.
  Emission of CO, C I, and C II in W3 Main. \emph{Astron. Astrophys.}
  \textbf{2004}, \emph{424}, 887--903\relax
\mciteBstWouldAddEndPuncttrue
\mciteSetBstMidEndSepPunct{\mcitedefaultmidpunct}
{\mcitedefaultendpunct}{\mcitedefaultseppunct}\relax
\EndOfBibitem
\bibitem[Anicich and Huntress(1986)Anicich, and Huntress]{Anicich:1986p38896}
Anicich,~V.~G.; Huntress,~W.~T. A Survey of Bimolecular Ion-molecule Reactions
  for Use in Modeling the Chemistry of Planetary Atmospheres, Cometary Comae,
  and Interstellar Clouds. \emph{Astrophys. J., Suppl. Ser.} \textbf{1986},
  \emph{62}, 553--672\relax
\mciteBstWouldAddEndPuncttrue
\mciteSetBstMidEndSepPunct{\mcitedefaultmidpunct}
{\mcitedefaultendpunct}{\mcitedefaultseppunct}\relax
\EndOfBibitem
\bibitem[Klippenstein et~al.(2010)Klippenstein, Georgievskii, and
  McCall]{Klippenstein:2010p41000}
Klippenstein,~S.~J.; Georgievskii,~Y.; McCall,~B.~J. Temperature Dependence of
  Two Key Interstellar Reactions of H$_3^+$: O($^3$P) + H$_3^+$ and CO +
  H$_3^+$. \emph{J. Phys. Chem. A} \textbf{2010}, \emph{114}, 278--290\relax
\mciteBstWouldAddEndPuncttrue
\mciteSetBstMidEndSepPunct{\mcitedefaultmidpunct}
{\mcitedefaultendpunct}{\mcitedefaultseppunct}\relax
\EndOfBibitem
\bibitem[Sofia et~al.(2004)Sofia, Lauroesch, Meyer, and
  Cartledge]{Sofia:2004p40486}
Sofia,~U.~J.; Lauroesch,~J.~T.; Meyer,~D.~M.; Cartledge,~S. I.~B. Interstellar
  Carbon in Translucent Sight Lines. \emph{Astrophys. J.} \textbf{2004},
  \emph{605}, 272--277\relax
\mciteBstWouldAddEndPuncttrue
\mciteSetBstMidEndSepPunct{\mcitedefaultmidpunct}
{\mcitedefaultendpunct}{\mcitedefaultseppunct}\relax
\EndOfBibitem
\bibitem[Lacy et~al.(1994)Lacy, Knacke, Geballe, and Tokunaga]{Lacy:1994p24447}
Lacy,~J.~H.; Knacke,~R.; Geballe,~T.~R.; Tokunaga,~A.~T. Detection of
  Absorption by H$_2$ in Molecular Clouds: A Direct Measurement of the H$_2$:CO
  Ratio. \emph{Astrophys. J.} \textbf{1994}, \emph{428}, L69--L72\relax
\mciteBstWouldAddEndPuncttrue
\mciteSetBstMidEndSepPunct{\mcitedefaultmidpunct}
{\mcitedefaultendpunct}{\mcitedefaultseppunct}\relax
\EndOfBibitem
\bibitem[Sodroski et~al.(1995)Sodroski, Odegard, Dwek, Hauser, Franz, Freedman,
  Kelsall, Wall, Berriman, Odenwald, and et~al.]{Sodroski:1995p40498}
Sodroski,~T.~J.; Odegard,~N.; Dwek,~E.; Hauser,~M.~G.; Franz,~B.~A.;
  Freedman,~I.; Kelsall,~T.; Wall,~W.~F.; Berriman,~G.~B.; Odenwald,~S.~F.;
  et~al., The Ratio of H$_2$ Column Density to $^{12}$CO Intensity in the
  Vicinity of the Galactic Center. \emph{Astrophys. J.} \textbf{1995},
  \emph{452}, 262--268\relax
\mciteBstWouldAddEndPuncttrue
\mciteSetBstMidEndSepPunct{\mcitedefaultmidpunct}
{\mcitedefaultendpunct}{\mcitedefaultseppunct}\relax
\EndOfBibitem
\bibitem[Arimoto et~al.(1996)Arimoto, Sofue, and Tsujimoto]{Arimoto:1996p40506}
Arimoto,~N.; Sofue,~Y.; Tsujimoto,~T. CO-to-H$_2$ Conversion Factor in
  Galaxies. \emph{Publ. Astron. Soc. Jpn.} \textbf{1996}, \emph{48},
  275--284\relax
\mciteBstWouldAddEndPuncttrue
\mciteSetBstMidEndSepPunct{\mcitedefaultmidpunct}
{\mcitedefaultendpunct}{\mcitedefaultseppunct}\relax
\EndOfBibitem
\bibitem[Rolleston et~al.(2000)Rolleston, Smartt, Dufton, and
  Ryans]{Rolleston:2000p40514}
Rolleston,~W. R.~J.; Smartt,~S.~J.; Dufton,~P.~L.; Ryans,~R. S.~I. The Galactic
  Metallicity Gradient. \emph{Astron. Astrophys.} \textbf{2000}, \emph{363},
  537--554\relax
\mciteBstWouldAddEndPuncttrue
\mciteSetBstMidEndSepPunct{\mcitedefaultmidpunct}
{\mcitedefaultendpunct}{\mcitedefaultseppunct}\relax
\EndOfBibitem
\bibitem[Chiappini et~al.(2001)Chiappini, Matteucci, and
  Romano]{Chiappini:2001p40528}
Chiappini,~C.; Matteucci,~F.; Romano,~D. Abundance Gradients and the Formation
  of the Milky Way. \emph{Astrophys. J.} \textbf{2001}, \emph{554},
  1044--1058\relax
\mciteBstWouldAddEndPuncttrue
\mciteSetBstMidEndSepPunct{\mcitedefaultmidpunct}
{\mcitedefaultendpunct}{\mcitedefaultseppunct}\relax
\EndOfBibitem
\bibitem[Esteban et~al.(2005)Esteban, Garc{\'\i}a-Rojas, Peimbert, Peimbert,
  Ruiz, Rodr{\'\i}guez, and Carigi]{Esteban:2005p40536}
Esteban,~C.; Garc{\'\i}a-Rojas,~J.; Peimbert,~M.; Peimbert,~A.; Ruiz,~M.~T.;
  Rodr{\'\i}guez,~M.; Carigi,~L. Carbon and Oxygen Galactic Gradients:
  Observational Values from H II Region Recombination Lines. \emph{Astrophys.
  J.} \textbf{2005}, \emph{618}, L95--L98\relax
\mciteBstWouldAddEndPuncttrue
\mciteSetBstMidEndSepPunct{\mcitedefaultmidpunct}
{\mcitedefaultendpunct}{\mcitedefaultseppunct}\relax
\EndOfBibitem
\bibitem[Wilms et~al.(2000)Wilms, Allen, and McCray]{Wilms:2000p43635}
Wilms,~J.; Allen,~A.; McCray,~R. On the Absorption of X-Rays in the
  Interstellar Medium. \emph{Astrophys. J.} \textbf{2000}, \emph{542},
  914--924\relax
\mciteBstWouldAddEndPuncttrue
\mciteSetBstMidEndSepPunct{\mcitedefaultmidpunct}
{\mcitedefaultendpunct}{\mcitedefaultseppunct}\relax
\EndOfBibitem
\bibitem[Aharonian et~al.(2004)Aharonian, Akhperjanian, Aye, Bazer-Bachi,
  Beilicke, Benbow, Berge, Berghaus, Bernl{\"o}hr, Bolz, and
  et~al.]{Aharonian:2004p38239}
Aharonian,~F.; Akhperjanian,~A.~G.; Aye,~K.-M.; Bazer-Bachi,~A.~R.;
  Beilicke,~M.; Benbow,~W.; Berge,~D.; Berghaus,~P.; Bernl{\"o}hr,~K.;
  Bolz,~O.; et~al., Very High Energy Gamma Rays from the Direction of
  Sagittarius A*. \emph{Astron. Astrophys.} \textbf{2004}, \emph{425},
  L13--L17\relax
\mciteBstWouldAddEndPuncttrue
\mciteSetBstMidEndSepPunct{\mcitedefaultmidpunct}
{\mcitedefaultendpunct}{\mcitedefaultseppunct}\relax
\EndOfBibitem
\bibitem[Aharonian and Neronov(2005)Aharonian, and
  Neronov]{Aharonian:2005p34402}
Aharonian,~F.; Neronov,~A. TeV Gamma Rays From the Galactic Center Direct and
  Indirect Links to the Massive Black Hole in Sgr A. \emph{Astrophys. Space
  Sci.} \textbf{2005}, \emph{300}, 255--265\relax
\mciteBstWouldAddEndPuncttrue
\mciteSetBstMidEndSepPunct{\mcitedefaultmidpunct}
{\mcitedefaultendpunct}{\mcitedefaultseppunct}\relax
\EndOfBibitem
\bibitem[Chernyakova et~al.(2011)Chernyakova, Malyshev, Aharonian, Crocker, and
  Jones]{Chernyakova:2011p34287}
Chernyakova,~M.; Malyshev,~D.; Aharonian,~F.~A.; Crocker,~R.~M.; Jones,~D.~I.
  The High-Energy, Arcminute-Scale Galactic Center Gamma-Ray Source.
  \emph{Astrophys. J.} \textbf{2011}, \emph{726}, 60\relax
\mciteBstWouldAddEndPuncttrue
\mciteSetBstMidEndSepPunct{\mcitedefaultmidpunct}
{\mcitedefaultendpunct}{\mcitedefaultseppunct}\relax
\EndOfBibitem
\bibitem[Bethe(1933)]{Bethe:1933p39189}
Bethe,~H. Handbuch der Physik: Aufbau der zusammenh{\"a}ngenden Materie.
  \emph{Handbuch der Physik} \textbf{1933}, \relax
\mciteBstWouldAddEndPunctfalse
\mciteSetBstMidEndSepPunct{\mcitedefaultmidpunct}
{}{\mcitedefaultseppunct}\relax
\EndOfBibitem
\bibitem[Webber(1998)]{Webber:1998p31383}
Webber,~W.~R. A New Estimate of the Local Interstellar Energy Density and
  Ionization Rate of Galactic Cosmic Cosmic Rays. \emph{Astrophys. J.}
  \textbf{1998}, \emph{506}, 329--334\relax
\mciteBstWouldAddEndPuncttrue
\mciteSetBstMidEndSepPunct{\mcitedefaultmidpunct}
{\mcitedefaultendpunct}{\mcitedefaultseppunct}\relax
\EndOfBibitem
\bibitem[Webber and Higbie(2009)Webber, and Higbie]{Webber:2009p39512}
Webber,~W.~R.; Higbie,~P.~R. Galactic Propagation of Cosmic Ray Nuclei in a
  Model with an Increasing Diffusion Coefficient at Low Rigidities: A
  Comparison of the New Interstellar Spectra with Voyager Data in the Outer
  Heliosphere. \emph{J. Geophys. Res.} \textbf{2009}, \emph{114}, A02103\relax
\mciteBstWouldAddEndPuncttrue
\mciteSetBstMidEndSepPunct{\mcitedefaultmidpunct}
{\mcitedefaultendpunct}{\mcitedefaultseppunct}\relax
\EndOfBibitem
\bibitem[Langner and Potgieter(2005)Langner, and Potgieter]{Langner:2005p40106}
Langner,~U.~W.; Potgieter,~M.~S. Modulation of Galactic Protons in an
  Asymmetrical Heliosphere. \emph{Astrophys. J.} \textbf{2005}, \emph{630},
  1114--1124\relax
\mciteBstWouldAddEndPuncttrue
\mciteSetBstMidEndSepPunct{\mcitedefaultmidpunct}
{\mcitedefaultendpunct}{\mcitedefaultseppunct}\relax
\EndOfBibitem
\bibitem[Scherer et~al.(2011)Scherer, Fichtner, Strauss, Ferreira, Potgieter,
  and Fahr]{Scherer:2011p40086}
Scherer,~K.; Fichtner,~H.; Strauss,~R.~D.; Ferreira,~S. E.~S.;
  Potgieter,~M.~S.; Fahr,~H.-J. On Cosmic Ray Modulation beyond the Heliopause:
  Where is the Modulation Boundary? \emph{Astrophys. J.} \textbf{2011},
  \emph{735}, 128\relax
\mciteBstWouldAddEndPuncttrue
\mciteSetBstMidEndSepPunct{\mcitedefaultmidpunct}
{\mcitedefaultendpunct}{\mcitedefaultseppunct}\relax
\EndOfBibitem
\bibitem[Moskalenko et~al.(2002)Moskalenko, Strong, Ormes, and
  Potgieter]{Moskalenko:2002p37919}
Moskalenko,~I.~V.; Strong,~A.~W.; Ormes,~J.~F.; Potgieter,~M.~S. Secondary
  Antiprotons and Propagation of Cosmic Rays in the Galaxy and Heliosphere.
  \emph{Astrophys. J.} \textbf{2002}, \emph{565}, 280--296\relax
\mciteBstWouldAddEndPuncttrue
\mciteSetBstMidEndSepPunct{\mcitedefaultmidpunct}
{\mcitedefaultendpunct}{\mcitedefaultseppunct}\relax
\EndOfBibitem
\bibitem[Cravens and Dalgarno(1978)Cravens, and Dalgarno]{Cravens:1978p38220}
Cravens,~T.~E.; Dalgarno,~A. Ionization, Dissociation, and Heating Efficiencies
  of Cosmic Rays in a Gas of Molecular Hydrogen. \emph{Astrophys. J.}
  \textbf{1978}, \emph{219}, 750--752\relax
\mciteBstWouldAddEndPuncttrue
\mciteSetBstMidEndSepPunct{\mcitedefaultmidpunct}
{\mcitedefaultendpunct}{\mcitedefaultseppunct}\relax
\EndOfBibitem
\bibitem[Neronov et~al.(2012)Neronov, Semikoz, and Taylor]{Neronov:2012p35437}
Neronov,~A.; Semikoz,~D.~V.; Taylor,~A.~M. Low-Energy Break in the Spectrum of
  Galactic Cosmic Rays. \emph{Phys. Rev. Lett.} \textbf{2012}, \emph{108},
  51105\relax
\mciteBstWouldAddEndPuncttrue
\mciteSetBstMidEndSepPunct{\mcitedefaultmidpunct}
{\mcitedefaultendpunct}{\mcitedefaultseppunct}\relax
\EndOfBibitem
\bibitem[Nath et~al.(2012)Nath, Gupta, and Biermann]{Nath:2012p34546}
Nath,~B.~B.; Gupta,~N.; Biermann,~P.~L. Spectrum and Ionization Rate of
  Low-Energy Galactic Cosmic Rays. \emph{Mon. Not. R. Astron. Soc.: Lett.}
  \textbf{2012}, \emph{425}, L86--L90\relax
\mciteBstWouldAddEndPuncttrue
\mciteSetBstMidEndSepPunct{\mcitedefaultmidpunct}
{\mcitedefaultendpunct}{\mcitedefaultseppunct}\relax
\EndOfBibitem
\bibitem[Liszt(2006)]{Liszt:2006p8116}
Liszt,~H.~S. H$_3^+$ in the Diffuse Interstellar Medium. \emph{Philos. Trans.
  R. Soc., A} \textbf{2006}, \emph{364}, 3049--3062\relax
\mciteBstWouldAddEndPuncttrue
\mciteSetBstMidEndSepPunct{\mcitedefaultmidpunct}
{\mcitedefaultendpunct}{\mcitedefaultseppunct}\relax
\EndOfBibitem
\bibitem[Liszt(2007)]{Liszt:2007p33674}
Liszt,~H.~S. Time-dependent H$_{2}$ Formation and Protonation in Diffuse
  Clouds. \emph{Astron. Astrophys.} \textbf{2007}, \emph{461}, 205--214\relax
\mciteBstWouldAddEndPuncttrue
\mciteSetBstMidEndSepPunct{\mcitedefaultmidpunct}
{\mcitedefaultendpunct}{\mcitedefaultseppunct}\relax
\EndOfBibitem
\bibitem[Genzel et~al.(2010)Genzel, Eisenhauer, and
  Gillessen]{Genzel:2010p37134}
Genzel,~R.; Eisenhauer,~F.; Gillessen,~S. The Galactic Center Massive Black
  Hole and Nuclear Star Cluster. \emph{Rev. Mod. Phys.} \textbf{2010},
  \emph{82}, 3121--3195\relax
\mciteBstWouldAddEndPuncttrue
\mciteSetBstMidEndSepPunct{\mcitedefaultmidpunct}
{\mcitedefaultendpunct}{\mcitedefaultseppunct}\relax
\EndOfBibitem
\bibitem[Yan et~al.(1998)Yan, Sadeghpour, and Dalgarno]{Yan:1998p37828}
Yan,~M.; Sadeghpour,~H.~R.; Dalgarno,~A. Photoionization Cross Sections of He
  and H$_2$. \emph{Astrophys. J.} \textbf{1998}, \emph{496}, 1044--1050\relax
\mciteBstWouldAddEndPuncttrue
\mciteSetBstMidEndSepPunct{\mcitedefaultmidpunct}
{\mcitedefaultendpunct}{\mcitedefaultseppunct}\relax
\EndOfBibitem
\bibitem[Yan et~al.(2001)Yan, Sadeghpour, and Dalgarno]{Yan:2001p37834}
Yan,~M.; Sadeghpour,~H.~R.; Dalgarno,~A. Erratum: Photoionization Cross
  Sections of He and H$_2$. \emph{Astrophys. J.} \textbf{2001}, \emph{559},
  1194\relax
\mciteBstWouldAddEndPuncttrue
\mciteSetBstMidEndSepPunct{\mcitedefaultmidpunct}
{\mcitedefaultendpunct}{\mcitedefaultseppunct}\relax
\EndOfBibitem
\bibitem[Hubbell et~al.(1975)Hubbell, Veigele, Briggs, Brown, Cromer, and
  Howerton]{Hubbell:1975p37862}
Hubbell,~J.~H.; Veigele,~W.~J.; Briggs,~E.~A.; Brown,~R.~T.; Cromer,~R.~T.;
  Howerton,~R.~J. Atomic Form Factors, Incoherent Scatteirng Functions, and
  Photon Scattering Cross Sections. \emph{J. Phys. Chem. Ref. Data}
  \textbf{1975}, \emph{4}, 471--538\relax
\mciteBstWouldAddEndPuncttrue
\mciteSetBstMidEndSepPunct{\mcitedefaultmidpunct}
{\mcitedefaultendpunct}{\mcitedefaultseppunct}\relax
\EndOfBibitem
\bibitem[Dalgarno et~al.(1999)Dalgarno, Yan, and Liu]{Dalgarno:1999p37811}
Dalgarno,~A.; Yan,~M.; Liu,~W. Electron Energy Deposition in a Gas Mixture of
  Atomic and Molecular Hydrogen and Helium. \emph{Astrophys. J., Suppl. Ser.}
  \textbf{1999}, \emph{125}, 237--256\relax
\mciteBstWouldAddEndPuncttrue
\mciteSetBstMidEndSepPunct{\mcitedefaultmidpunct}
{\mcitedefaultendpunct}{\mcitedefaultseppunct}\relax
\EndOfBibitem
\bibitem[Meijerink and Spaans(2005)Meijerink, and Spaans]{Meijerink:2005p36531}
Meijerink,~R.; Spaans,~M. Diagnostics of Irradiated Gas in Galaxy Nuclei. I. A
  Far-Ultraviolet and X-Ray Dominated Region Code. \emph{Astron. Astrophys.}
  \textbf{2005}, \emph{436}, 397--409\relax
\mciteBstWouldAddEndPuncttrue
\mciteSetBstMidEndSepPunct{\mcitedefaultmidpunct}
{\mcitedefaultendpunct}{\mcitedefaultseppunct}\relax
\EndOfBibitem
\bibitem[Maillard et~al.(2004)Maillard, Paumard, Stolovy, and
  Rigaut]{Maillard:2004p38227}
Maillard,~J.~P.; Paumard,~T.; Stolovy,~S.~R.; Rigaut,~F. The Nature of the
  Galactic Center Source IRS 13 Revealed by High Spatial Resolution in the
  Infrared. \emph{Astron. Astrophys.} \textbf{2004}, \emph{423}, 155--167\relax
\mciteBstWouldAddEndPuncttrue
\mciteSetBstMidEndSepPunct{\mcitedefaultmidpunct}
{\mcitedefaultendpunct}{\mcitedefaultseppunct}\relax
\EndOfBibitem
\bibitem[Raymond and Smith(1977)Raymond, and Smith]{Raymond:1977p38228}
Raymond,~J.~C.; Smith,~B.~W. Soft X-ray Spectrum of a Hot Plasma.
  \emph{Astrophys. J., Suppl. Ser.} \textbf{1977}, \emph{35}, 419--439\relax
\mciteBstWouldAddEndPuncttrue
\mciteSetBstMidEndSepPunct{\mcitedefaultmidpunct}
{\mcitedefaultendpunct}{\mcitedefaultseppunct}\relax
\EndOfBibitem
\bibitem[Foster et~al.(2012)Foster, Ji, Smith, and
  Brickhouse]{Foster:2012p38116}
Foster,~A.~R.; Ji,~L.; Smith,~R.~K.; Brickhouse,~N.~S. Updated Atomic Data and
  Calculations for X-Ray Spectroscopy. \emph{Astrophys. J.} \textbf{2012},
  \emph{756}, 128\relax
\mciteBstWouldAddEndPuncttrue
\mciteSetBstMidEndSepPunct{\mcitedefaultmidpunct}
{\mcitedefaultendpunct}{\mcitedefaultseppunct}\relax
\EndOfBibitem
\bibitem[Rockefeller et~al.(2004)Rockefeller, Fryer, Melia, and
  Warren]{Rockefeller:2004p36525}
Rockefeller,~G.; Fryer,~C.~L.; Melia,~F.; Warren,~M.~S. Diffuse X-Rays from the
  Inner 3 Parsecs of the Galaxy. \emph{Astrophys. J.} \textbf{2004},
  \emph{604}, 662--670\relax
\mciteBstWouldAddEndPuncttrue
\mciteSetBstMidEndSepPunct{\mcitedefaultmidpunct}
{\mcitedefaultendpunct}{\mcitedefaultseppunct}\relax
\EndOfBibitem
\bibitem[Koyama et~al.(1996)Koyama, Maeda, Sonobe, Takeshima, Tanaka, and
  Yamauchi]{Koyama:1996p39395}
Koyama,~K.; Maeda,~Y.; Sonobe,~T.; Takeshima,~T.; Tanaka,~Y.; Yamauchi,~S. ASCA
  View of Our Galactic Center: Remains of Past Activities in X-Rays?
  \emph{Publ. Astron. Soc. Jpn.} \textbf{1996}, \emph{48}, 249--255\relax
\mciteBstWouldAddEndPuncttrue
\mciteSetBstMidEndSepPunct{\mcitedefaultmidpunct}
{\mcitedefaultendpunct}{\mcitedefaultseppunct}\relax
\EndOfBibitem
\end{mcitethebibliography}



\end{document}